\documentclass[twocolumn,showpacs,amssymb,aps]{revtex4}

\usepackage{graphicx}
\usepackage{epsfig}
\usepackage{dcolumn}
\usepackage{bm}

\def\lessim{\lower.5ex\hbox{$\; \buildrel < \over \sim \;$}}
\begin{document} \hbadness=10000
\topmargin -0.8cm\oddsidemargin = -0.7cm\evensidemargin = -0.7cm
\preprint{}

\title{Strangeness Chemical Equilibration in QGP at RHIC and LHC}
\author{Jean Letessier}
\affiliation{Laboratoire de Physique Th\'eorique et Hautes Energies\\
Universit\'e Paris 7, 2 place Jussieu, F--75251 Cedex 05
}
\author{Johann Rafelski}
\affiliation{Department of Physics, University of Arizona, Tucson, Arizona, 85721, USA}

\date{July 30, 2006}

\begin{abstract}
We study,  in the   dynamically evolving QGP fireball formed in relativistic heavy ion collisions 
at RHIC and LHC, the growth of strangeness yield toward and beyond the chemical equilibrium.
 We account for the  contribution  of the direct strangeness production 
and evaluate  the thermal-QCD   strangeness production mechanisms. The 
specific yield of strangeness per entropy,$s/S$, is the primary target variable.   We explore 
the effect of collision impact parameter, {\it i.e.}, fireball size, on kinetic strangeness 
chemical equilibration in QGP.   Insights gained in study  the RHIC data with regard to the
dynamics of the fireball are applied to the  study  strangeness production at the LHC. 
We use these results and consider the strange hadron relative  particle yields at RHIC and LHC
in a systematic fashion. We consider both the dependence on $s/S$ and  directly participant number
dependence.   
\end{abstract}

\pacs{25.75.Nq, 25.75.-q, 24.10.Pa,12.38.Mh}
 \maketitle

 
\section{Introduction}\label{Intro}
Conversion of kinetic collision energy into high multiplicity of newly made hadronic 
particles is one of the most notable  features of reactions observed at the Relativistic
Heavy Ion Collider (RHIC) at the Brookhaven National Laboratory (BNL)~\cite{Back:2004je}. 
 In this process, aside of the light $u$ and $d$ quark pairs,  present in all matter surrounding us, 
the strange flavor quark pairs  $s,\bar s$  are produced copiously  (in general, in what follows, 
the particle symbol will refer to the corresponding  particle yield, either total or per unit of 
rapidity, as appropriate). The final  $s$-yield depends  on the initial reactions, and 
on the history of the fireball, and thus, also on the  nature and properties
 of the phase of matter formed. 
On the time scale of hadronic interactions, strangeness flavor is conserved, and 
prior to any weak interaction decays,  we have  $\bar s=s$ --- when we refer to strangeness
yield, production, etc, we always address  yield, production, etc, of strange quark pairs.   

This study is  addressing strangeness under the physical conditions 
achieved, at RHIC, at the highest attainable reaction
energy today, $\sqrt{s_{\rm NN}}=200$ GeV, and in future at LHC.
We are particularly interested in the sensitivity of strangeness production
to the nature and properties of the matter formed in the heavy ion reactions. 
Theoretical studies have shown that strangeness is produced rapidly
in collisions (fusion) 
of thermalized gluons~\cite{Rafelski:1982pu,Letessier:1996ad}, 
within the deconfined state,
the quark--gluon plasma (QGP) formed in the central collisions of 
heavy nuclei. 
On the scale of RHIC reaction time $\tau<10$ fm, the
hadron based reactions  were found 
to be  too slow to allow copious 
strangeness production after thermalization 
of matter  and  are even more ineffective 
to produce multi-strange hadrons~\cite{Koch:1984tz}.  

On the other hand, strangeness can be produced fast in the QGP phase, as we shall see 
achieving near chemical equilibrium in QGP phase formed at RHIC and even overshooting 
the chemical equilibrium at hadronization at 
the LHC. Thus,  the situation is quite different when the deconfined QGP state breaks up
in a fast hadronization process: the   enhancement of strange hadrons  and most specifically
strange antibaryons,  growing 
with valance strange quark content of hadrons produced is the 
 predicted   characteristic property of the 
deconfined QGP phase~\cite{Koch:1986ud}. This happens since
in the breakup of the strangeness rich deconfined state,  {\it i.e.}, hadronization, several
strange quarks formed in prior, and independent, reactions can combine into a multistrange hadron.

Our main objective in this work is to quantify the 
mechanisms of kinetic strangeness production
occurring in thermal gluon collision (fusion) processes 
during the expansion phase of the quark--gluon fireball. 
 We explore the centrality dependence 
in a wide  range between peripheral, and most central reactions,
in which up to 90\% of projectile and target  nucleons participate.  In this regard,
this paper is  a  theoretical companion to the more phenomenological analysis of 
experimental RHIC data~\cite{Rafelski:2004dp,Letessier:2005kc}, and uses the insights gained 
in this analysis, in particular, regarding the dynamics of  the QGP expansion. This is then
applied to extrapolate our approach to the LHC energy domain, where our prior particle 
yield study was based on a parametric consideration of final state strangeness 
yield~\cite{Rafelski:2005jc}.

We also study  in depth the centrality dependence of strangeness
production in the RHIC-LHC energy range. As the centrality of the nuclear reaction 
and the number of participants $A$ decreases, the number of thermal 
collisions gradually diminishes, and  with it the strangeness enhancement 
effect also diminishes gradually, similarly to the AGS-SPS energy range~\cite{Letessier:1996ad,ActaB}.
This decrease drives in turn a gradual decrease in the centrality dependent 
production rate of multistrange hadrons.
This behavior we discuss in detail here 
is an important and characteristic phenomenological feature 
of the kinetic particle collision  mechanism 
of strangeness production and enhancement. 

Interestingly, in this regard the 
kinetic mechanism of strangeness production  differs from 
models, such as `canonical enhancement model', which are deriving the strange hadron 
enhancement as a result of an always prevailing hadronic phase chemical  
equilibrium~\cite{Koch:1982ij,Braun-Munzinger:2003zd}. The volume 
dependence of  the   canonical  phase space  yields~\cite{Rafelski:1980gk}, and the
smallness of the N--N reference systems produce the centrality
 enhancement effect~\cite{Hamieh:2000tk}. However,  `canonical enhancement'  is rising very 
rapidly considering rather small collision systems, and with decreasing energy~\cite{Redlich:2001kb}. 

The understanding of strangeness production
during the expansion phase of the quark--gluon fireball allows us to 
study in depth the reaction mechanisms  which  are determining
the final state yield of strangeness. In this way, we learn how
`deep' into the history of QGP expansion this observable allows us 
to look.  Collective matter flow features observed
at RHIC suggest that thermalization of parton matter occurred
very fast, {\it i.e.}, the entropy $S$ has been produced in a not yet fully understood 
fast process of parton thermalization,  prior to the  production 
of strangeness pair yield $s$.   We formulate the 
kinetic equations allowing to address,  in  some detail, the growth in specific strangeness 
per (fixed) entropy $s/S$ in  the thermal QGP processes in  both a longitudinally and 
transversely  expanding QGP fireball. 

The time  evolution of 
$s/S$ has been  considered for the first time early on in the
development of the QGP physics~\cite{Kapusta:1986cb}. However, the model of 
dense matter evolution,  and the range of statistical parameters considered at the time 
is not appropriate for the RHIC and LHC physics environments. Moreover, we recompute here
the rate of strangeness production knowing the best current values of QCD parameters,
the coupling constant $\alpha_s$, and the strange quark mass $m_s$. These parameters
alter decisively the values of $s/S$ in chemical equilibrium, and thus the dynamical time
scale of the approach to chemical equilibrium. In order to 
be able to compute the evolution in time of $s/S$, 
we must also evaluate how near to chemical equilibrium  is
strangeness in QGP: this nearness is characterized by a parameter $\gamma_s$, roughly 
the ratio of prevailing strangeness density to chemical equilibrium density at
prevailing temperature.  

Setting up the production of strangeness, we assume here that it   follows in time the  
chemical equilibration of the light $q=u,\,d$ quarks,  and $g$ gluons~\cite{Alam:1994sc},
which are believed to occur at 1 fm scale.  These effectively massless particles 
can be produced by entirely soft processes which are intrinsically 
non-perturbative~\cite{Geiger:1992si}. Their chemical equilibration can 
be further driven by multi-particle
collisions~\cite{Xiong:1992cu,Xu:2004mz}. 
Considering these studies, we  assume here relatively short relaxation times for 
$q,\,g$, we cannot  compute these using the same perturbative method as will be developed 
here to evaluate  strangeness yield equilibration. It is the finite strangeness mass combined
with the  measured strength of the running QCD coupling constant $\alpha_s$ which allows 
us the use of perturbative formalism in study of strangeness 
production~\cite{Rafelski:2001kc} with some minimal confidence. 

The chemical relaxation times for the 
strangeness  approach to chemical equilibrium, in an expanding QGP, 
has been considered several times
before~\cite{Koch:1986hf,Matsui:1985eu,Biro:1993qt,Letessier:1996ad,Rafelski:1999gq,Pal:2001fz,He:2004df}.
The study of QGP strangeness chemical equilibration must not be confused with the 
phenomenological investigation of   chemical equilibrium in the final state hadron 
abundance. Because hadron phase space is generally smaller, chemical equilibrium and indeed
excess over equilibrium is much more easy to attain, and there is some   continuing
discussion of this question~\cite{Letessier:1998ca,Becattini:2003wp,Torrieri:2005va}. 

We offer in our work a 
comprehensive exploration how the impact parameter dependence, and 
consideration of energy dependence, influences chemical equilibration in QGP. 
 We evolve in time not the strangeness  itself, but the specific strangeness 
per entropy $s/S$. In this way, we can identify  
more clearly the production processes of strangeness,
since entropy is produced earlier on, and is (nearly) conserved
during the time period of thermal strangeness production. 

Importantly,  $s/S$ is an experimental observable, practically 
preserved in the fast hadronization process. 
Thus, we can connect the final state of the 
QGP evolution directly to experimental soft hadron yield experimental 
results.   We find that much of the variability about
the initial conditions, such as dependence on initial temperature, cancels in the 
final result. This specific observable will be shown to yield nearly model independent 
insights about thermal strangeness production in QGP. 

In the following section \ref{over}, we  begin  with a brief discussion of 
general features relevant in all considerations presented. 
 In subsection \ref{thermal}, we formulate the kinetic equations describing   the 
growth in specific strangeness  per entropy $s/S$ and show 
that, at RHIC, the  observed specific per entropy 
strangeness yield suggests that the 
`direct' and `thermal' processes are of comparable strength in 
most central  reactions.  We discuss the magnitude and importance
of QCD parameters in subsection \ref{QCDinput}.
In order to integrate as function of reaction time the strangeness yield equations,
we develop a simple collective expansion model of the plasma phase
in subsection \ref{expand}. 

In section \ref{ress}, we study the thermal strangeness  production processes.
We then obtain  reference yields for two 
different expansion geometries at RHIC in subsection \ref{bench}. We
extrapolate this  to the LHC environment in subsection \ref{LHC}.
 We explore  how `deep' into the early history of the
hot and dense fireball the strangeness signature of thermal QGP 
is allowing us to look, {\it i.e.}, the dependence on initial 
conditions, in subsection \ref{initial} -- importantly we find that the selection of the initial 
value of $s/S$ related to direct production of strangeness, has only minor impact on the
final results regarding strangeness yield.  We then explore the influence of fundamental 
uncertainties, such as  the present day 
limited knowledge about the strange quark mass and 
the QCD thermal effects on the freezing of the strange quark degrees of freedom,
in subsection  \ref{Fund}. 

In section \ref{eval}, we connect the results we obtained to the experimental particle 
yields. In subsection \ref{sh}, we   consider  the relationship between hadron multiplicity
and entropy yield, and obtain the strange hadron yield as function of $s/S$. This
allows an assessment how strangeness yield, at RHIC and LHC, influence physical observables.
In particular, we discuss how  K$^+/\pi^+$  changes between these two experimental 
environments. Then, we
discuss, in subsection \ref{PY}, for the two most often used statistical hadronization models
(sudden hadronization and chemical equilibrium hadronic gas (HG) hadronization) the production
of strange hadrons as function of participant number $A$, keeping the hadronization 
condition independent of $A$. In subsection \ref{thc}, 
we apply the insights gained in study of thermal strangeness production  to evaluate 
thermal charm production at the RHIC and the LHC environments.

\section{Remarks about Strangeness Production and Density }\label{over}
\subsection{Parton equilibration and strangeness production}

The total final state hadron multiplicity is a  measure of the entropy $S$ produced  
prior to thermal production of strangeness $s$ in QGP: 
once a quasi-thermal exponential energy  distribution of partons 
has been formed, the entropy production has been completed. 
Further evolution of the dense deconfined fireball is nearly 
entropy conserving, even though it is  strangeness flavor producing: 
fusion of gluons, or light quark pair annihilation into strangeness, is 
a nearly entropy conserving process~\cite{Elze:2001ss}.  We note  that, 
in reactions between two thermal particles into a strangeness pair, the energy
content of each initial state parton is transferred to  the two reaction products,
so thermal partons produce thermal shape of  strangeness spectrum.

The entropy produced, in RHIC reactions, has been evaluated in recent studies of 
hadron multiplicities. In Au--Au reactions, at $\sqrt{s_{\rm NN}}=200$ GeV, 
one sees $S\simeq 35,000$. Furthermore, at central rapidity, the yield
of entropy is $dS/dy\simeq 5000$. In the benchmark results we present
for LHC, we will assume that the the central
rapidity entropy yield is about 4 times greater than at RHIC. 

The temporal evolution of the QGP fireball ends  
 when the temperature has decreased to the 
QGP hadronization value. In the breakup of the QGP, the 
yields of hadrons are established, and it is rather difficult in
the ensuing rather short lived evolution lasting not more than 
1.5 fm/$c$ to alter these yields appreciably. Thus, 
the hadronization volume, with the normalizing factor  $dV/dy$,
provides the normalization of hadron particle 
yields per unit of rapidity. The final value of  $dV(\tau_f)/dy$   is result of 
 analysis of hadron particle yields and our model of the time
dependence of $dV(\tau)/dy$ will be constrained by the  
magnitude of   $dV_f/dy$ obtained in Ref.~\cite{Rafelski:2004dp}.

There are two separate stages of 
strangeness (charm) production, corresponding to the two practically distinct periods
of the fireball evolution:\\
a) `direct' production  
creates a `background' yield corresponding to what might be obtained in 
a superposition model of independent nucleon--nucleon (N--N) reactions. \\
b) strangeness production in  collisions between thermally equilibrated QGP
constituents. 

For b) to be relevant   prior formation of a thermal,
deconfined  QGP phase is required. Without process b),
the yield of strangeness should not be enhanced in A--A collisions 
as compared to scaled N--N reactions. We will not discuss in detail mechanisms a) 
of direct particle production here, our interest is  restricted to
(approximate) initial   yields, which are the baseline for the thermal 
mechanisms acting in the QGP, and define the strength of any 
enhancement. 

We note for the record, that 
strangeness initial production, like other relatively soft parton
production processes, are believed to be due to 
color string breaking mechanism~\cite{Sjostrand:2003wg} and the 
Pythia 6 model of soft hadron production presumes that the relative
strength of $u:d:s$ production is $1:1:0.3$. The
(initial) charm production is   due to 
high energy parton collisions~\cite{Bedjidian:2003gd}. 

Our study of  the thermal strange particle production processes 
is based on kinetic theory of particle collisions.  
There is considerable uncertainty about the initial   momentum
distributions of soft partons present in the initial state. However,
two recent theoretical studies argue that there is rapid thermalization. 
The  nonlinear gluon production
processes leads to the  gluon momentum
distribution equilibration~\cite{Xu:2004mz}.  
Axial asymmetry of the  initial state causes collective
instabilities which further accelerate thermalization
of partons~\cite{Mrowczynski:2005ki}. 

In this context, it is important to advance one result of our study, namely that 
there is little sensitivity   to the initial
thermal condition: a wide range of `reasonable' initial temperatures 
leading to very similar strangeness production 
results, as long as the entropy content is 
preserved. In order to understand this, consider  a 
decrease in initial temperature. This requires, at fixed entropy, an increase
in initial volume, and this, then, is associated with increased lifespan
of the fireball   in the high temperature strangeness producing domain. These 
two effects combine to compensate the reduced strangeness production 
rate per unit of time and  volume that is associated with reduced 
ambient temperature. 

We believe that this mechanism 
also implies that the   precise form of the momentum
spectrum of the initial state partons is of minor practical 
relevance for the purpose of evaluation of strangeness production,
and we do not study this. 
Therefore, without  loss of generality, we can  assume that the parton 
distributions we use in the kinetic  strangeness formation process have
thermal shape, and the ambient temperature is determined considering 
(lattice fitted) equations of state relation of initial temperature 
and entropy density~\cite{Letessier:2003uj},  for  a given geometric initial volume. 

We already remarked above that the production of strangeness,
 in a cascade of N--N reactions (without deconfinement), is 
not able to add significantly to the initial strangeness yield considering the 
short lifespan of the   fireball. Thus, this alternative will not
be further considered in this work. Similarly, any additional strangeness produced in 
the rapid hadronization of QGP into hadrons must be negligible compared to the 
thermal production process which occurs at higher particle density (temperature), and 
during a considerably longer lifespan. 

\subsection{Approach to chemical yield equilibrium}\label{chemapp}
It has been shown, considering the entropy maximization principle,
that the  approach of particle densities to chemical 
equilibrium  density $\rho_i^\infty$ can be characterized by the statistical parameter
$\gamma_i$~\cite{Letessier:1993qa}, which varies with the local proper time $\tau$ 
during the collision. For example, for gluons,
\begin{equation}\label{gamma}
\rho_{\rm g}(\tau)\equiv 
 \int d^3p \frac{\gamma_{\rm g}(\tau) e^{-E/T}}{1-\gamma_{\rm g}(\tau) e^{-E/T}}, 
\quad E=\sqrt{m^2+p^2}.
\end{equation}
Generally, the Lagrangian mass of gluons is zero. However, one may
be tempted to think that thermal mass $m(T)$ could change decisively 
results, suppressing the collisional strangeness production. However,
one finds that instead the process of gluon decay becomes relevant
and if at all, there is a net rate increase of strangeness production~\cite{Biro:1990vj}.
One could argue that the scheme to study kinetic process of strangeness chemical 
equilibration using 
thermal mass amounts to a different resummation of reaction processes. 
In this work,   we will consider the evolution of $\gamma_{s}^{\rm QGP}(\tau)$ 
based on Lagrangian masses, allowing for $\gamma_{\rm g}^{\rm QGP}(\tau) $
and $\gamma_{q}^{\rm QGP}(\tau) $. 

For strange quarks, we will keep only the Boltzmann term 
(ignoring the denominator in Eq.\,(\ref{gamma}) above) and thus:
\begin{equation}\label{sdens}
\rho_s(\tau)\equiv \gamma_s^{\rm QGP}(\tau)\rho_s^\infty 
  =\gamma_s^{\rm QGP}\,{g_s\over 2\pi^2}z^2K_2(z),\  z=\frac{m_s}{ T}.
\end{equation}
Here, $g_s$ is the strange quark degeneracy,   and $K_2$ is a Bessel function.
Since we will  employ strangeness occupancy in hadron phase,  $\gamma_s^{\rm h}$,
we have included the subscript  {\small QGP}. 
We will henceforth drop this subscript, and occupancy parameters without
superscript will, in general, refer to the  QGP phase, while the hadronic gas 
phase variable, when these are expected to differ from QGP, will have a superscript $h$. 

Our target variable is the final QGP state specific yield of strangeness per entropy, 
$s/S$, and the related phase space occupancy $\gamma_s$. Both these 
variables have an important
physical relevance: $s/S$ determines the  final yield of strange hadrons 
compared to all hadrons, and its value implies some particular yield of 
reference yields, such as, {\it e.g.}, K$^+/\pi^+$.   $\gamma_s$ characterizes the 
approach to chemical equilibrium, it measures 
 strangeness yield in terms of the chemical equilibrium yield. 
The strangeness phase space of QGP and HG phases are different.
Strangeness in QGP is much denser than in the HG phase,
considering the range of strange quark masses, $0.080<m_s(\mu=2\, {\rm GeV})<0.125$ GeV.
Therefore,  (near) chemical strangeness
equilibrium  in the QGP phase  $\gamma_s\simeq 1$,  implies   
a significantly oversaturated hadron phase space abundance 
$\gamma_s^{\rm h}>1$ after hadronization. $\gamma_s^{\rm h}$ is directly 
controlling the relative yields of hadrons with different $s+\bar s$
valance quark content and is thus observable. 

\subsection{Role of initial conditions}\label{Rini}

We do not understand well the conditions in the QGP phase at time as early as
$\tau_0=0.25$ fm for RHIC, and $\tau_0=0.1$ fm 
for LHC, when we presume that the thermal momentum
distribution is practically established. Thus, we must make 
a number of  assumptions and check if these impact our results.
The relevant parameters that could govern the strangeness production are:\\
$\gamma_{\rm g}(\tau)$ and, in particular, the initial value at $\tau_0$;\\
$\tau_0$, the time at which we assume thermal momentum of partons is reached;\\
$s/S|_{\tau_0}$ is the initial strangeness yield originating in direct parton collisions; \\
$R_\bot$ is the   transverse radius dimension at initial time,   related to the collision 
geometry;\\
$v_\bot(\tau)$ is the transverse  expansion velocity, and
 in particular its maximum value at hadronization; \\
$\tau_{\rm g}$ and $\tau_q$,   the  relaxation time constant of   gluon and quark fugacities,
considering that quarks are less relevant compared to gluons with regard 
to strangeness production, we will assume $\tau_q$=1.5 $\tau_{\rm g}$ throughout this work.

We explore, here, a characteristic gluon thermalization
time $0.1< \tau_0 <1.5$ fm/$c$,  with the longest
value applicable to most peripheral RHIC Au--Au reactions at $\sqrt{s_{\rm NN}}=200$ GeV,
and the shortest period assumed for the future LHC 
central collisions. Knowing the exact  dynamics of thermalization and how
long it takes will be, as we shall see, rather unimportant for the final insights we
obtain. 

All initial state  parameters are  constrained in their value, either by
collision geometry,  by final state particle yields observed at RHIC, or/and 
by particle  correlations.  For example, the final yield of strangeness $ds/dy$
and of entropy $dS/dy$, and thus $s/S$ are known from an analysis of  
particle production:   as function of 
centrality    at RHIC~\cite{Rafelski:2004dp}, and as function of reaction energy
from top AGS, SPS to top RHIC energy~\cite{Letessier:2005qe}.

We will use these results in two ways. We compute,  following the  temporal
evolution,  the final state $s/S$ ratio which we expect to converge at RHIC to 
$s/S\simeq 0.033$. We need to specify the initial value at time $\tau_0$ for
this variable, and this value is chosen to be compatible with the peripheral 
reactions. The entropy yield $dS/dy$, which we assume is 
conserved during the evolution of QGP, determines, for a 
known initial volume $dV(\tau=\tau_0)/dy$,  the entropy density $\sigma=(dS/dy)/(dV/dy)$.
We then can obtain, from standard properties of QGP fitted to the lattice
results, the initial temperature $T_0/T_c\simeq 3$--4. This temperature decreases
as volume expands with $\tau$ given that the entropy is preserved.

We will show
that the two physical observables, $s/S$ and $\gamma_s$, we address
are largely independent of the model dependent details
of the initial conditions. Said differently, our important finding is 
that the two global  strangeness observables  $s/S$ and $\gamma_s$ appear
to penetrate back only to about 2 fm/$c$  after the reaction has begun, and do not probe 
earlier conditions in the QGP phase. The physical reason for this is,
of course, that once chemical 
equilibrium is approached,   one looses much of the event memory with regard to 
intensive physical observables. We will further see that,  in cases we studied, 
that did not quite reach chemical equilibrium in the 
QGP phase, this is also true, {\it i.e.}, there is little sensitivity 
to what exactly happened to light quarks and gluons
 in the first 2 fm/s. 

The reason for this  is that there is a strong correlation 
between volume, temperature and degree of QGP (gluon and light quark)
chemical equilibration as we already discussed above. Repeating the argument
differently,  we can say that when  fewer gluons at fixed 
entropy are in given volume, temperature has to be larger. Thus, any 
decrease in the production of strangeness in gluon fusion due to absence of 
gluons is compensated by the greater specific rate per colliding pair due
to greater ambient temperature. Hence, also when we do not quite reach chemical
equilibrium in QGP, be it due to large impact parameter or low reaction 
energy (chemical nonequilibrium at lower energies is 
not explored in this paper), there is little if any dependence 
of thermal yield on initial conditions, and  the results we arrive at  regarding
near chemical equilibration are extraordinarily robust. 

However,   the  thermal strangeness production does depend 
on the degree of initial state strangeness equilibration, 
simply because, if the initial yields were chemically equilibrated to start with,
there would be as much production as annihilation of strangeness and any
temporal evolution is driven by the time dependence of the evolution dynamics. 
We will take as a measure of the pre-QGP thermal
 phase   strangeness production the specific 
per hadron multiplicity yield  of strangeness observed in 
most peripheral RHIC reactions. 
This is typically $s/S\simeq 0.016$ at $\tau=\tau_0$. This choice 
allows  to reproduce 
the observed value $s/S=0.019$ attained in most peripheral nuclear
reactions at RHIC~\cite{Rafelski:2004dp}, 
with participant number about $\langle A\rangle =6.3$.
Clearly, with $s/S\to 0.033$ in most central collisions, the implication
of this choice is that the thermal process enhances total specific 
yield by   factor $1.9\pm 0.3$. As centrality and/or reaction energy 
decreases, there is a gradual decrease of this enhancement, and
as the energy is increased (LHC) this enhancement rises somewhat.

\subsection{Strangeness production in thermal collisions}\label{thermal}
We follow   the   established
methods of evaluating thermal strangeness production~\cite{Letessier:2002gp}, expanding
our earlier more schematic model~\cite{Rafelski:1999gq}.
However, considerable simplification arises since we
focus attention on the specific yield of strangeness per entropy.

In the local (comoving) frame of reference, the rate 
of change of strangeness   is due to production 
and annihilation reactions only:
\begin{eqnarray} 
\frac{1}{V} {{d s}\over {d \tau}}=\frac{1}{V}{{d  {\bar s}}\over {d \tau} }
&=& 
\frac12 \rho_{\rm g}^2(t)\,\langle\sigma v \rangle_T^{gg\to s\bar s}
+
\rho_{q}(t)\rho_{\bar q}(t)
\langle\sigma  \rangle_T^{q\bar q\to s\bar s}\nonumber \\ \label{qprod}
&&- 
\rho_{s}(t)\,\rho_{\bar{\rm s}}(t)\,
\langle\sigma v\rangle_T^{s\bar s\to gg,q\bar q}.
\end{eqnarray} 
The thermally average cross sections
are: 
\begin{equation}\label{Tsig}
\langle\sigma v_{\rm rel}\rangle_T\equiv
\frac{\int d^3p_1\int d^3p_2 \sigma_{12} v_{12}f(\vec p_1,T)f(\vec p_2,T)}
{\int d^3p_1\int d^3p_2 f(\vec p_1,T)f(\vec p_2,T)}\,.
\end{equation}
$f(\vec p_i,T)$ are the relativistic Boltzmann/J\"uttner
distributions of two colliding particles $i=1,2$ of momentum $p_i$,
characterized by local statistical parameters. 

 A convenient way to address the  dilution  phenomena acting on the 
density of strangeness $\rho_s\equiv s/V$ due to rapid expansion of
the QGP phase, Eq.\,(\ref{qprod}), is to  consider the proper time evolution 
of the specific strangeness per entropy yield:
\begin{equation}\label{qprod3}
{d\over d\tau} {s\over S}=\frac VS \  \frac1V\ {d {s}\over d \tau}. 
\end{equation}
The entropy $S$
in a volume  element is unchanged,  as volume grows and temperature drops:
\begin{equation}\label{S1}
S=V{4\pi^2\over 90} g(T)T^3={\rm Const.},
\end{equation}
where we consider the quark and gluon degrees of freedom along with their
QCD corrections:
\begin{eqnarray}\label{ggq}
g&=&2_s8_c\left(1-\frac{ 15\alpha_s(T)}{4\pi}+\ldots\right)\nonumber\\ 
 &&+\frac74 2_s3_cn_{\rm f}  \left(1-\frac{ 50\alpha_s(T)}{21\pi}+\ldots\right).
\end{eqnarray}
We use as the number of quark flavors
$n_{\rm f}\simeq 2+ \gamma_s0.5 z^2\,K_2(z)$, where $z=m_s/T$.
The   terms  proportional to chemical potentials
are not shown in the expression for entropy, since $\mu/\pi T\ll 1$ at RHIC and LHC. 

We have used, here, the lowest order  QCD corrections to the effective degeneracies,
since these describe well the properties of QGP 
phase obtained on the lattice~\cite{Letessier:2003uj}, when
the value of $\alpha_s(T)$ used is as described below, see Eq.\,(\ref{alfaseq}). 
The agreement one sees for thermodynamic variables, such as $E,\,P,\,S$,
 with the lattice results is very remarkable, including the temperature range near
to the phase  boundary. Thus, the use of constraint Eq.\,(\ref{S1})
to evaluate the time dependence of temperature, considering also that the
third root of entropy is considered, should yield precise enough results.
We note, in passing, that we used as specified in Ref.~\cite{Letessier:2003uj}
the additional terms ${\cal A}$ in entropy, arising from differentiation 
of the implicit temperature dependence of $g$, Eq.\,(\ref{ggq}) entering the 
partition function.

In order to use the detailed balance
which relates production and annihilation reactions, 
it is convenient to introduce the invariant rate per unit time and volume, 
$A^{12\to 34}$, by incorporating the equilibrium densities into the thermally
averaged cross sections:
\begin{equation}
A^{12\to 34}\equiv\frac1{1+\delta_{1,2}}
\gamma_1\gamma_2  \rho_1^\infty\rho_2^\infty 
             \langle \sigma_{s} v_{12} \rangle_T^{12\to 34}.  
\end{equation}
$\delta_{1,2}=1$ for the reacting particles being identical bosons, 
and otherwise, $\delta_{1,2}=0$.  Note also that
the evolution for $s$ and $\bar s$ in proper time of the 
comoving volume element is identical as both change in pairs.

We find that the temporal evolution of $s/S$, in an expanding plasma, 
is governed by:  
\begin{eqnarray}\label{qprod3a}
 {d\over { d\tau}} {s\over S}
&=&
 {A^{gg\to s\bar s}\over (S/V) } 
    \left[\gamma_{\rm g}^2(\tau)-\gamma_{s}^2(\tau)\right]  \nonumber\\ 
&&+
 {A^{q\bar q\to s\bar s}\over (S/V) } 
    \left[\gamma_q^2(\tau)-\gamma_{ s}^2(\tau)\right]\,.
\end{eqnarray}
When all $\gamma_i\to 1$, the Boltzmann collision term vanishes, and
 equilibrium has been reached. The value arrived at for the observable $s/S$ depends
on the history of how the system evolves and, eventually, reaches equilibrium.

In order to be able to solve  Eq.\,(\ref{qprod3a}), we need a relation 
between $s/S$ and   $ \gamma_{ s}$. This is obtained combining strangeness 
density Eq.\,(\ref{sdens}) and entropy Eq.\,(\ref{S1}):
\begin{equation}\label{sS1}
\frac sS=\gamma_s  {g_s \over g} \frac{90}{8\pi^4}z^2K_2(z),\quad z=m_s/T.
\end{equation}
 In the initial period,  gluons and quarks have not reached chemical equilibrium,
thus the actual
numerical integrals of Bose and Fermi distributions of the type   Eq.\,(\ref{gamma}),
dependent on the values $\gamma_{q,{\rm g}}$ are employed  instead,  which modifies
the result seen in  Eq.\,(\ref{sS1}).

The degeneracies we have considered in Eq.\,(\ref{S1}) 
for the entropy did include the effect of interactions, and 
thus, we have to allow for  the interaction effect in the 
strange quark degeneracy as well:
\begin{equation}\label{gs}
g_s=2_s3_c\left(1-\frac{ k\alpha_s(T)}{\pi}+\ldots\right).
\end{equation}
The value of $k=2$ applies to massless strange quarks. At $T=0$ 
(or said differently, for $m\gg T$) the early study of quark matter 
self-energy  suggests that $k\to 0$~\cite{Chin:1979yb}. We will 
present, in figure \ref{alfdep} below, results for varying the 
value of  $k$, and the reference value we use in our other studies  
is $k=1$. We believe that this approach  allows us to explore the 
general behavior of the   interactions effect on the strangeness 
density, a more detailed study is not possible today.

\subsection{QCD parameters} \label{QCDinput}
We evaluate $A^{gg\to s\bar s}$ and $A^{q\bar q\to s\bar s}$ employing the available
strength of the QCD coupling, and range of accepted strange 
quark masses. In our evaluation of strangeness production,  in order
   to account for higher order effects in quark and gluon fusion
   reactions, we introduce a multiplicative  $K=1.7$-factor.
The known properties of QCD strongly constrain our 
results~\cite{Letessier:1996ad}. However, it turns out that the range
of strange quark masses remains sufficiently wide to impact the results
and we discuss this further in subsection~\ref{Fund}, see figure \ref{massdep} below. 
We employ when not otherwise stated the central value from a recent PDG evaluation,
$m_s(\mu=2\,{\rm GeV})=0.10$ GeV which remains uncertain at the level of 25\%
at least. In fact, since our results were obtained, a more recent Particle Data Group study 
recommends a 10\% smaller central value of $m_s$~\cite{Yao:2006px}.

We compute rate of reactions employing a running strange quark mass working
in two loops, and using as the energy scale the CM-reaction energy 
$\mu\simeq  \sqrt{s}$. Since the running of mass involves
a multiplicative factor, the uncertainty in the mass value discussed above
is  the same for all values of $\mu$. Some simplification is further
achieved by taking at temperature $T$ the value $\mu\simeq 2\pi T$ which 
is the preferred value of the thermal field theory, and   agrees with
 the value of the reaction energy. This means that we use $m_s(T)=m_s(\mu=2\pi T)$
with $m_s(T=318\,{\rm MeV})=0.1\,$GeV. The actual temperature  $T(t)$ and thus time 
dependent  values of the strange (and charm) quark mas will be always shown in the 
top panel of figures describing the evolution of the properties of the system
considered. 
 
\begin{figure}[t]
\hspace*{.2cm}
\psfig{width=8.5cm, figure=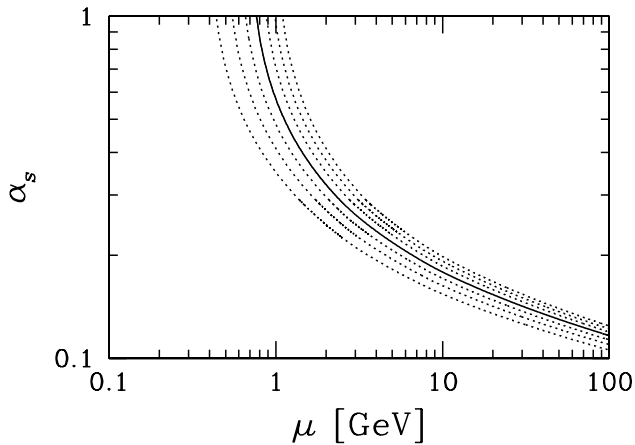}
\caption{\label{alfas}
The running QCD couping constant $\alpha_s(\mu)$ 
fixed to  $\alpha_s(\mu=m_{Z^0})=0.118$
(solid line) and several alternative strength scenarios excluded today 
by the experimental measurement (dashed lines). 
}
\end{figure}

The  strength  of  the QCD couping constant 
is today much better understood. We use  as reference value  
 $\alpha_s(\mu=m_{Z^0})=0.118$, 
and evolve the value to applicable energy domain $\mu$ by using two loops. 
Our ability to use perturbative methods of QCD to describe strangeness production,
a relatively soft process, 
derives from two circumstances:\\
a) the reaction processes which change yield of strangeness 
can compete with the fast $v_\bot>0.5c$ expansion of   QGP only for 
$T>220$ MeV, for lower temperatures the strange quark yields effectively
do not change (strange quark chemical freeze-out temperature in QGP). 
Using the relation $\mu=2\pi T$, this implies that all
strangeness yield evolution occurs for   $ \mu > 1.4 $\,GeV.\\
b) Because of the magnitude   $\alpha_s(\mu=m_{Z^0})=0.118$, one can 
quite well run  $\alpha_s$ to the scale of interest, $\mu>1.2GeV$.\\
As we see in figure \ref{alfas}, this means that the strength of the interaction 
remains $\alpha_s<0.5$. We also note that had the strength of $\alpha_s(\mu=m_{Z^0})$
been 15\% greater, strangeness production could not be studied in perturbative 
approach.

We next express 
$\alpha_s(\mu)$ (solid line in figure \ref{alfas}) as function of
temperature by the conditions $\alpha_s(T)=\alpha_s(\mu=2\pi T)$. 
This leads to the  expression (see also section 14 in~\cite{Letessier:2002gp}):
\begin{equation}\label{alfaseq}
\alpha_s(T)\simeq {\alpha_s(T_c)\over 1+C\ln (T/T_c)},\quad T<6 T_c,
\end{equation}
with $C=0.760\pm0.002$, $\alpha_s(T_c)=0.50\pm0.04$ at $T_c=0.16 $ GeV.
We stress that Eq.\,(\ref{alfaseq})  is a parametrization
corresponding to the result shown in figure \ref{alfas}. Only one logarithm
needs to be   used to describe the two loop
running with sufficient precision, since the range we consider
is rather limited, $0.9T_c<T<6T_c$.  
 
\subsection{Expansion and cooling of QGP} \label{expand}
We separate, in our work, the issue of strangeness production from
the even more complex and less understood questions about the 
time evolution of the QGP. We assume that there is some active
volume at average temperature $T$, in which the strangeness is `cooked'. 
We derive the time dependence of local temperature from the 
hypothesis of a conserved entropy content and a reasonable model 
describing the volume evolution in time. This is arrived at using
a hydrodynamically inspired model. 

The volume at hadronization is
an implicit observable. All particle yields  at hadronization are normalized 
with a volume factor. Thus, our expansion model must be realistic
enough so that the hadronization conditions are in agreement
with data, and that the impact parameter dependence is 
reproduced.  In the geometry inspired model we consider, in
 the central rapidity domain:
\begin{equation}\label{volmod}
{dV\over dy}= A_\bot(\tau)  \left.{dz\over dy}\right\vert_{\tau={\rm Const.}}.
\end{equation} 
$dV/dy$ is the normalization factor for the  particle yields we are measuring  
 in an interval around central rapidity. The transverse expansion is 
described by the transverse size $A_\bot(\tau)$.  We further need to associate with 
the domain of  rapidity $dy$  a geometric region at the 
source $dz$, from which particles emerge.

To accomplish this, we recall the space-time rapidity of
the scaling Bj\o rken hydrodynamical solution:
\begin{equation}\label{ybj}
y=\frac12 \ln {t+z\over t-z} . 
\end{equation}
We see that $y$=0 corresponds to $z=0$.
In particular, if the transverse extend of the fireball  is large, the Bj\o rken
space-time rapidity relation prevails. 

We need this relation   not at 
a fixed laboratory time $t$ but at some fixed proper time  $\tau$:
\begin{equation}\label{tau}
\tau=\sqrt{t^2- z^2 }.
\end{equation}
We eliminate in Eq.\,(\ref{ybj}) $t$ using Eq.\,(\ref{tau}):
\begin{equation}\label{zy}
z=\tau \sinh y,\quad {dz\over dy} = \tau \cosh y .
\end{equation}
 $A_\bot$ is the transverse to scattering axes size of the evolving
QGP. For nearly homogeneous expanding bulk matter one can  assume:
\begin{equation}
A_\bot =\pi R_\bot^2(\tau).
\end{equation} 
However, if the matter is predominantly concentrated near
a narrow   domain of width $d$, we
consider:
\begin{eqnarray}
A_\bot 
 &=&\pi \left[R_\bot^2(\tau)-(R_\bot^2(\tau)-d)^2\right],\nonumber\\
 &=&2\pi d \left[R_\bot (\tau)-\frac d2\right] .\label{Adonut}
\end{eqnarray}

At central rapidity, we consider quantitatively the two 
evolution scenarios, denoted here-forth as models V1 and V2. V1 will be
the most simple bulk homogeneous expansion while  V2 simulates a
transverse donut, it corresponds to expansion with a cold 
hole of matter in fireball (axial) center:
\begin{eqnarray}\label{V1}
&&{\bf\rm V1:}\quad {dV\over dy}= \pi R^2_\bot(\tau) \tau,\\
\label{V2}
&&{\bf\rm V2:}\quad  {dV\over dy}= 2\pi d \left[R_\bot (\tau)-\frac d2\right]  \tau,  
\end{eqnarray} 
with 
\begin{equation}
R_\bot(\tau)=R_0+\int v(\tau) d\tau .
\end{equation}
 
Any model  of transverse matter   expansion dynamics $v(\tau)$
is  constrained by the transverse mass shape of produced particle spectra,
too large transverse velocities would produce too hard spectra. 
Accordingly, a hydro-inspired shape is assumed: 
\begin{equation}\label{vexp}
v(\tau)= v_{\rm max} {2\over \pi} \arctan [4 (\tau-\tau_0)/\tau_v].
\end{equation}
Values of $v_{\rm max}$ we consider are  in the range of 0.5--0.8$c$, the relaxation time
$\tau_c\simeq 0.5$ fm, and the onset of transverse expansion $\tau_0$
was tried in range 0.1--1 fm. None of these parameters matters for what 
follows as long as one does not employ aberrant values.
 
The initial size $R_\bot$  is  assumed, in what follows, 
to be $R_\bot=5$\,fm for 5\% most central
 collisions. When we study centrality dependence, 
we  will show results 
for a series of centralities decreasing the transverse dimensions 
$R_\bot$ and $d$ by factor $f_R=1.5$ in each step. We further scale  entropy value
with  $f_S=f_R^{2.2}$. This assures that
the dependence  of entropy on the participant number $dS(A)/dy$ in
the final state follows the   relationship,  
\begin{equation}\label{entsca}
\frac{dS}{dy}\simeq 8 (A^{1.1}-1),
\end{equation}
 obtained from the impact parameter dependent fit  
to the RHIC impact parameter results~\cite{Rafelski:2004dp}.

\begin{figure}[t]
\vspace*{-0.3cm}
\psfig{width=8cm,height=8.5cm,figure=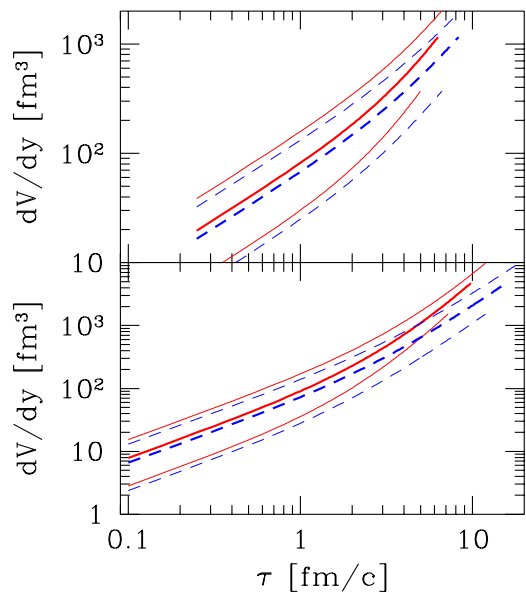}
\vspace*{-0.6cm}
\caption{\label{Volume}
(Color online) QGP Volume related to central rapidity, $dV/dy$ 
as function of   proper time $\tau$. 
Top panel is for RHIC 
with reference entropy content $dS/dy=5,000$ (central lines), 
while bottom panel is for LHC with
4 times greater entropy content $dS/dy=20,000$ (central lines).   
Three centralities are considered, with the middle 
thicker lines corresponding to   $R_\bot=5$ fm and 
the upper/lower lines corresponding to  $R_\bot=7 $, and, 
respectively,   $R_\bot=3 $ fm/$c$.
Solid lines are for 
V1 model with transverse homogeneity, dashed lines for 
V2 model of a transverse  shell with   widths (top to bottom)
 $d=2.1,\ 3.5$ and 4.9 fm.  
 The volume expansion is shown in
the figure up to   $T=140$ MeV. See text for more details. 
}
\end{figure}

An overview of the resulting volume dynamic 
behavior is given in figure \ref{Volume}, the top
panel applies to RHIC with $\sqrt{s_{\rm NN}}=200$ GeV, 
the bottom panel presents  a parallel study for LHC with the 
{\it assumed} four times greater entropy content. The 
solid lines are for transverse homogeneous volume (V1 model) 
expansion, and dashed lines correspond 
to a  transverse region of thickness $d=3.5$\,fm (V2 model). 
 
Three different centralities 
were considered with $R_\bot=3$, 5 and 7  fm. For the second model of transverse
expansion,  the transverse size $d$ is scaled with 
$R_\bot/5$ fm. Thus, $d=2.1$\,fm for $R_\bot=3$\,fm and 
 $d=4.9$\,fm for $R_\bot=7$\,fm.  Similarly, 
entropy content, assumed to be $dS/dy= 5000$ at RHIC and  $dS/dy=20,000$ at LHC
for $R_\bot=5$ fm, is scaled to values $dS(R_\bot=3 \,{\rm fm})/dy=1300$ and 
$dS(R_\bot=7 \,{\rm fm})/dy=10,500 $, and correspondingly, 
4 times greater values for LHC. 

The temporal expansion of the volume is followed till  
$T=140$ MeV is reached. In general, the maximum  volume  at LHC is
thus 4 times greater compared to RHIC. The expansion time is correspondingly longer,
with RHIC taking 6.5fm  to freeze-out for $R_\bot=5$fm, the LHC lifespan is 10fm. 
The QGP lifespan increase by as much as 60\%   at LHC, when comparing to RHIC,  if
the assumed initial entropy production is indeed increased by factor~4.  
  From the perspective of strangeness production, this 
is one of the more interesting changes comparing RHIC to LHC.

Given the volume  as function of $\tau$ and the associated 
conserved entropy content, we can evaluate the prevailing  temperature 
$T$ for any given  quark and gluon  chemical  yield condition
$\gamma_{q,s,{\rm g}}$.  The solid lines in top panels 
of the following figures \ref{TwoVol} to \ref{massdep} show this result, in the figures
  \ref{TwoVol} and \ref{LHCVolVol2},  on 
left for the   V1 model and on right for the   V2 (donut) model. 
The assumed $\gamma_{\rm g}$ is presented as dashed line.
In some of the top panels, we also show by the dotted line 
the time dependence of the applied transverse 
velocity, $v_\bot$, see Eq.\,(\ref{vexp}).

\section{Results on strangeness production}\label{ress}

\subsection{The benchmark results for RHIC}\label{bench}
\begin{figure*}[t]
\vskip -0.5cm
\psfig{width=7.9cm,figure=  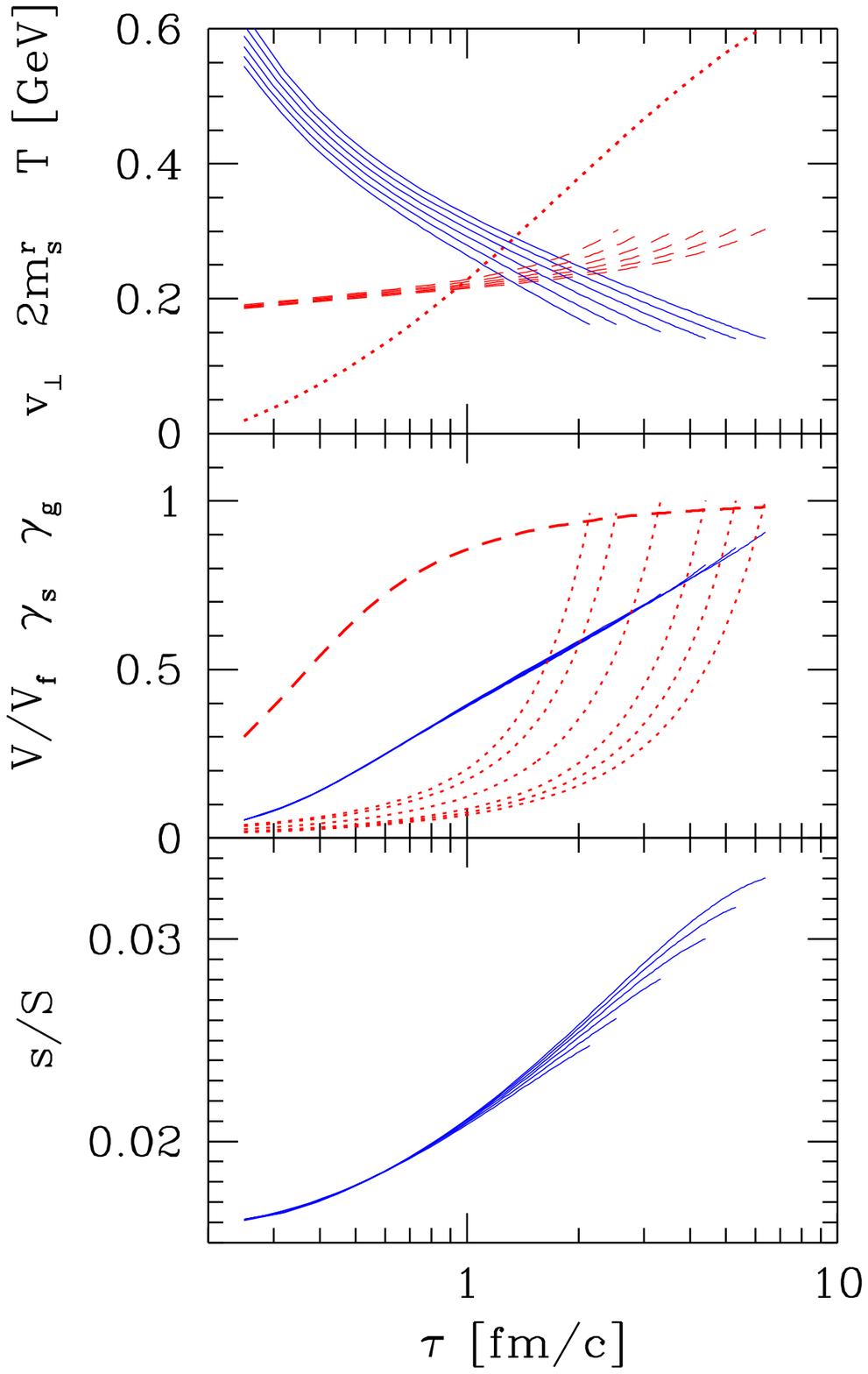  }
\psfig{width=7.9cm,figure=  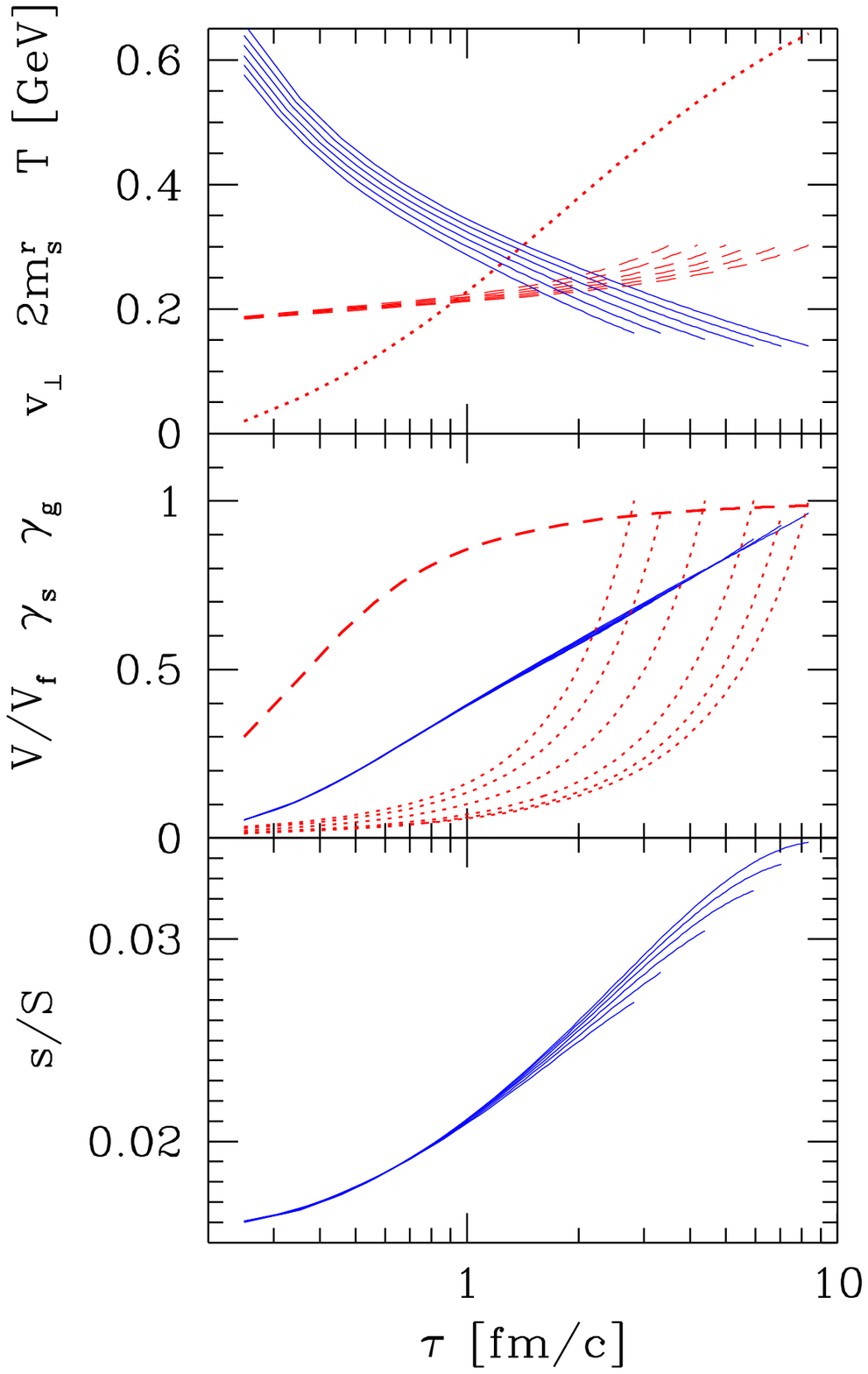    }
\vspace*{-0.6cm}
\caption{\label{TwoVol}
(Color online)  RHIC results.
Top panel: solid lines: temperature $T$; dashed lines: running mass $m_s^r(T)$;
 dotted line: the  assumed  profile of transverse expansion velocity 
$v_\bot(\tau)$. Different lines correspond 
to different centralities. 
Middle panel:  Solid line(s)  $\gamma_s$, which nearly coincide 
for different centralities; dashed, the assumed  $\gamma_{\rm g}(\tau)$, 
dotted the profile of the assumed  volume, 
$[dV(\tau)/dy]/[dV(\tau_f)/dy]$ normalized by the freeze-out value. 
$R_\bot(\tau_0)$ stepped down for each line by factor 1.5. 
The end points, at maximum
$\tau$, allow to identify corresponding  $s/S$ for different centrality in the 
bottom panel. 
Right and left: Comparison of the two transverse expansion models, 
see Eqs.\,(\ref{V1},\ref{V2}): left bulk 
expansion (model V1), right donut expansion (model V2).  
}
\end{figure*}

\begin{figure*}[t]
\vskip -0.5cm
\psfig{width=7.9cm,figure=  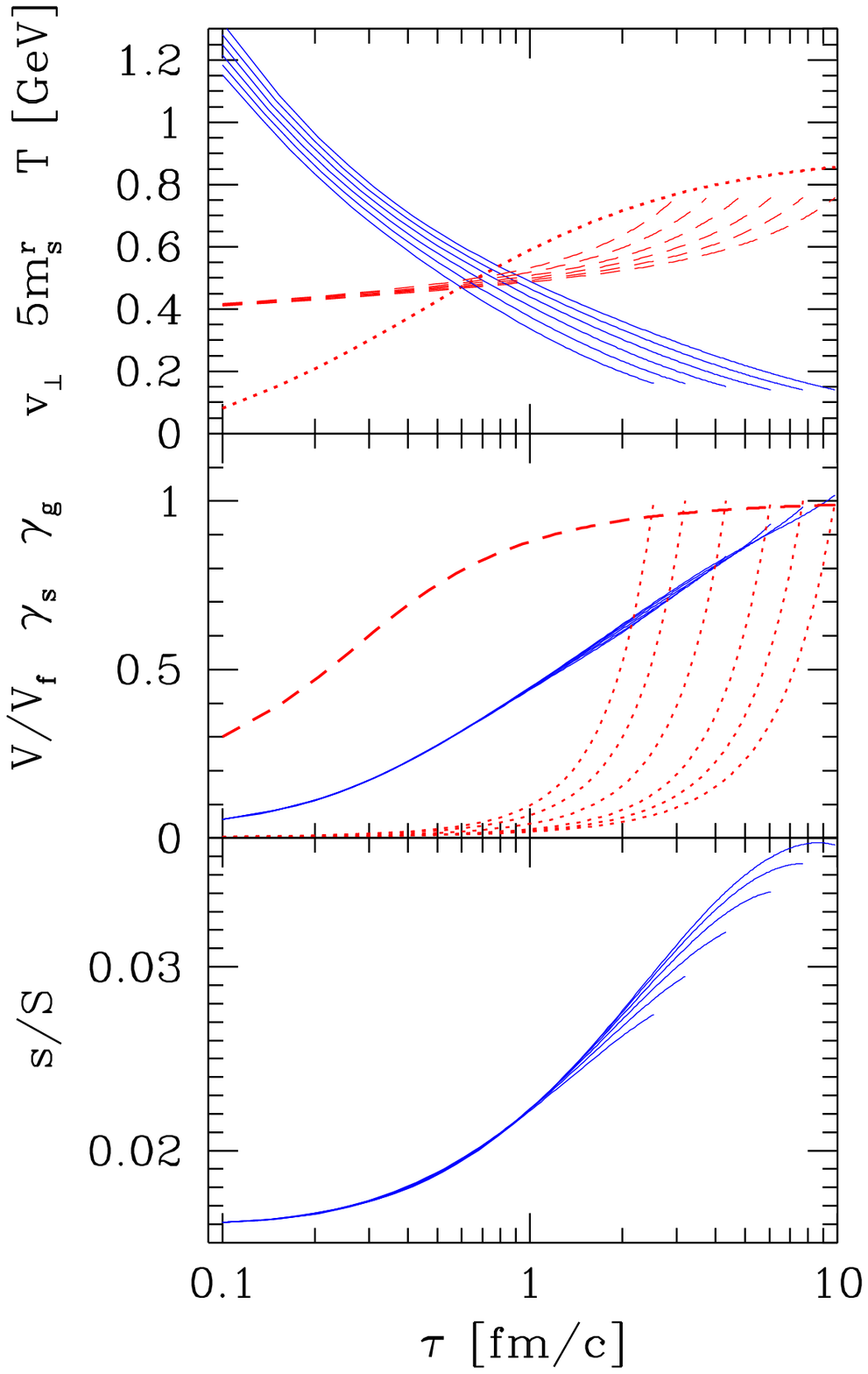  }
\psfig{width=7.9cm,figure=  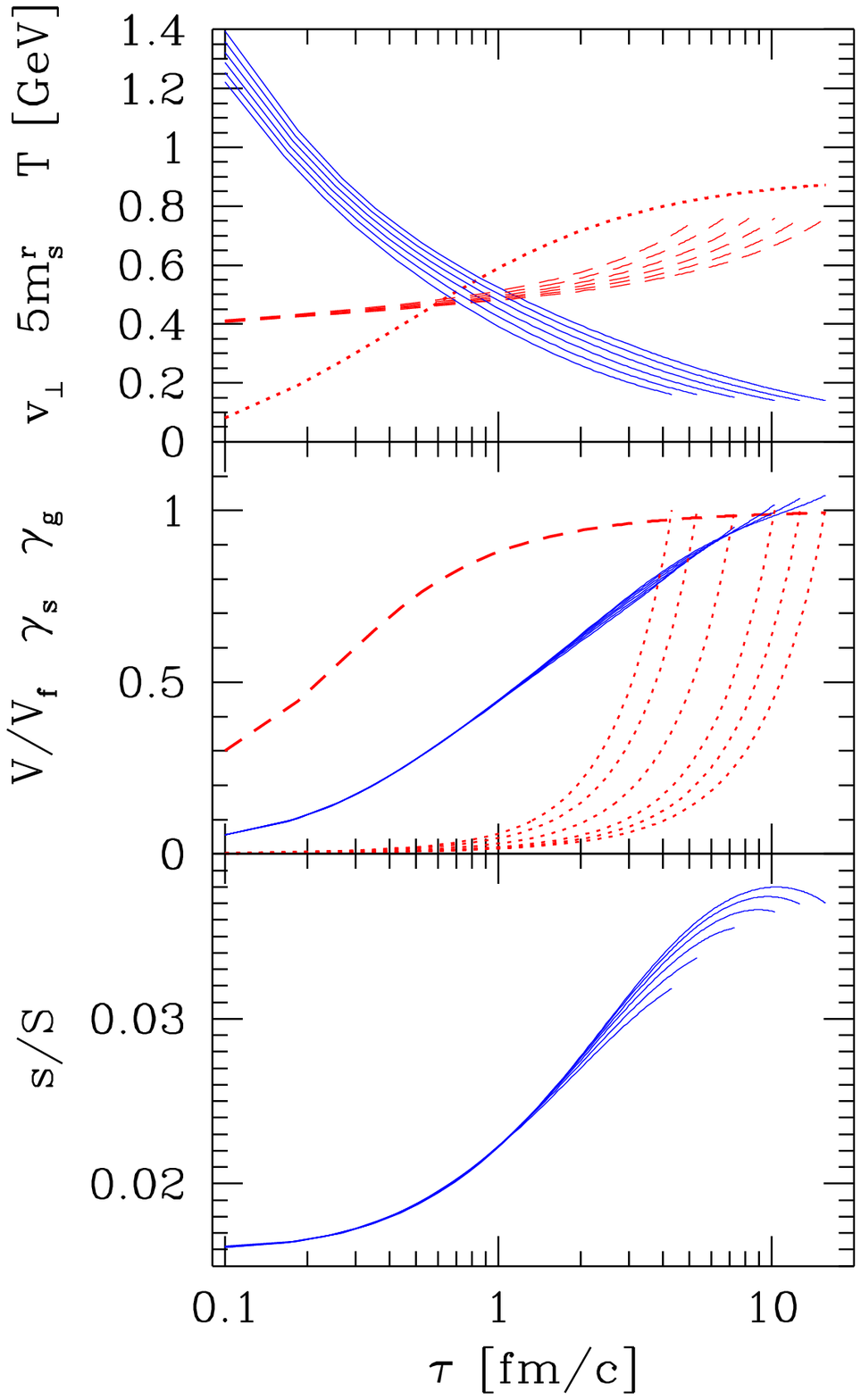   }
\vspace*{-0.6cm}
\caption{\label{LHCVolVol2}
(color online) Case of LHC, see legend in  figure \ref{TwoVol}. 
See text for discussion of  differences with RHIC.
}
\end{figure*}

\begin{figure*}[t]
\vskip -0.5cm
\psfig{width=7.9cm,figure=   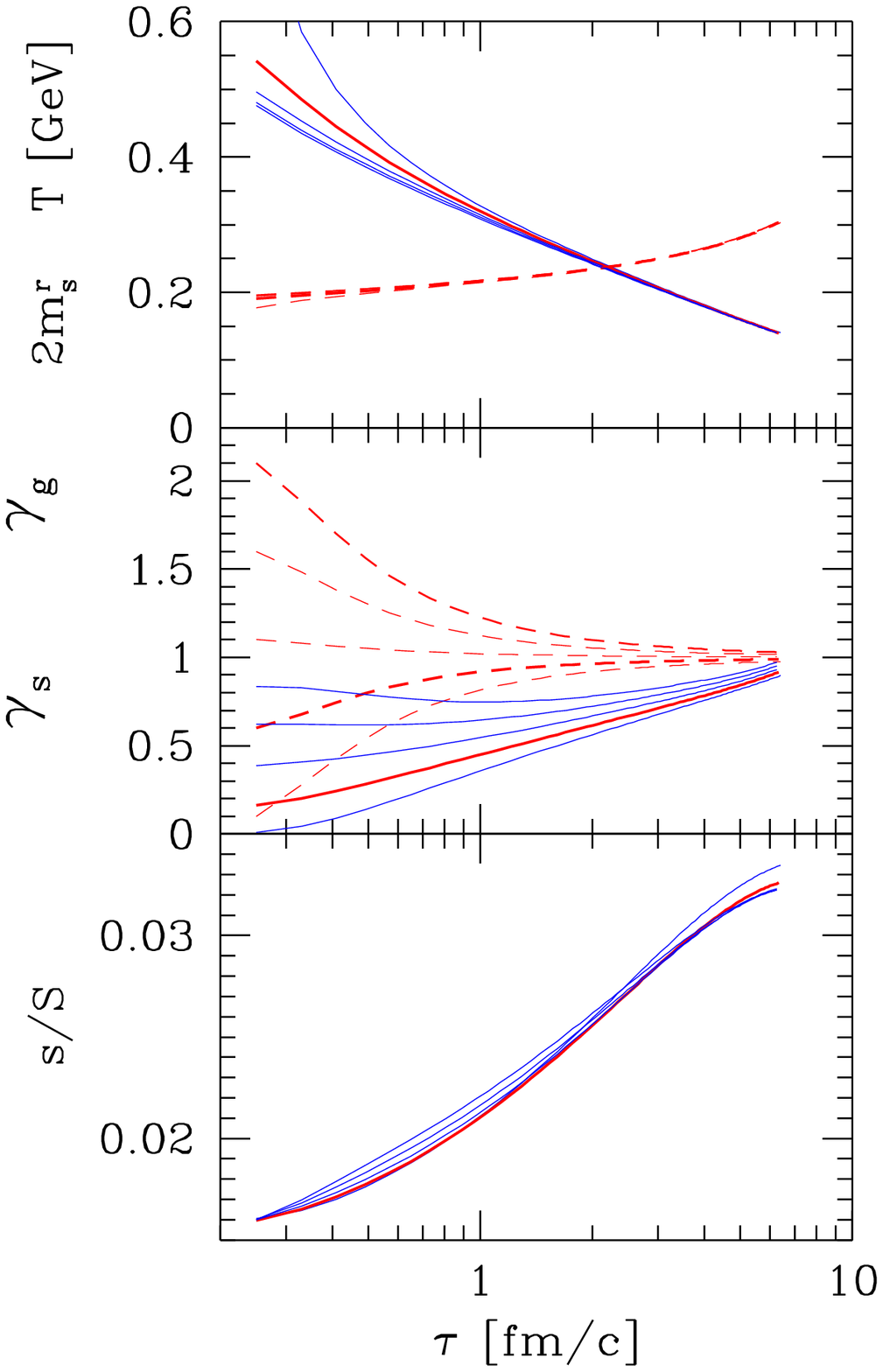  }
\psfig{width=7.9cm,figure= 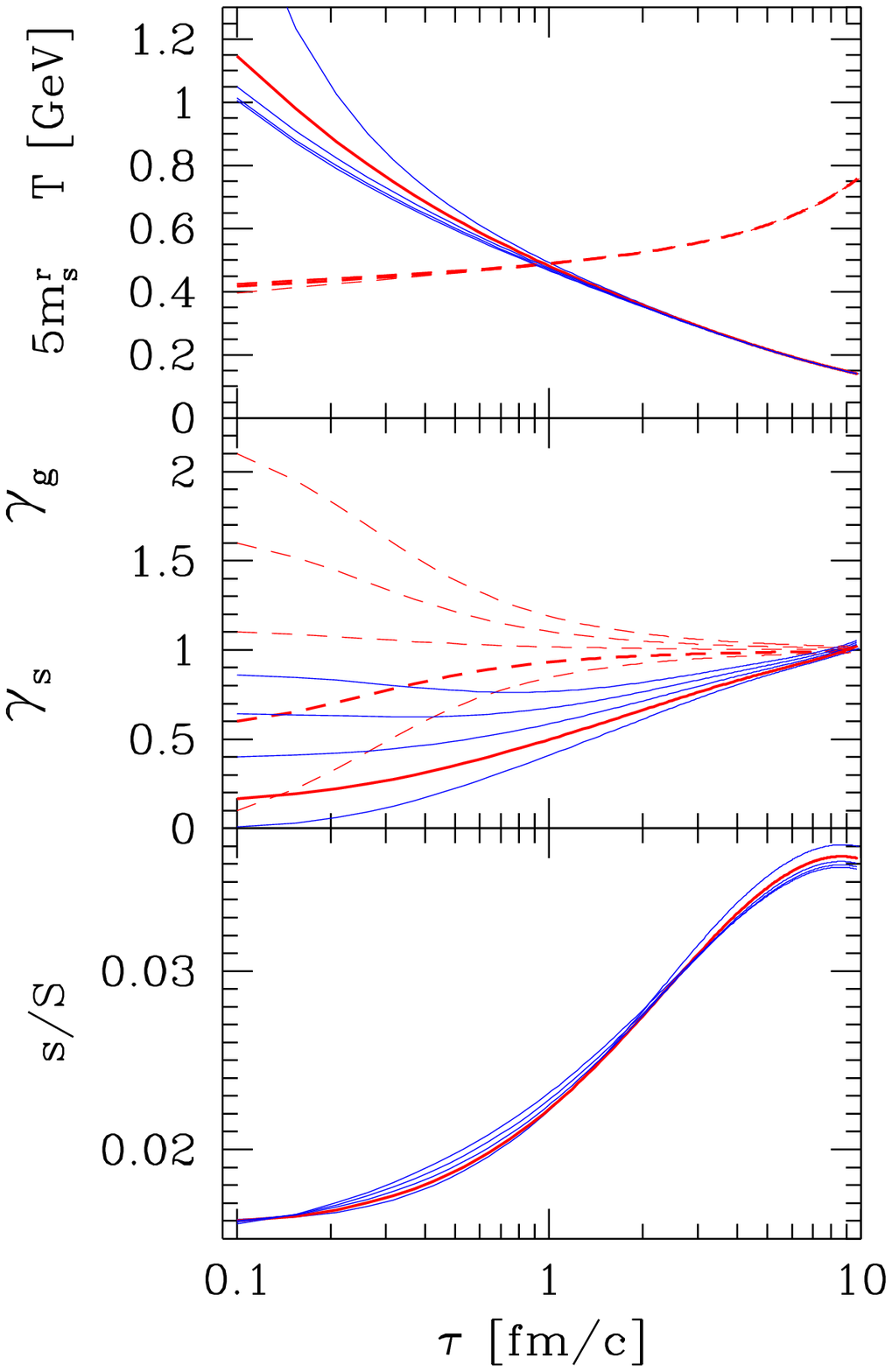   }
\vspace*{-0.6cm}
\caption{\label{Gluedep}
(color online) Model V1 (volume expansion) at RHIC (left) and LHC (right) 
for 5\% most central collisions. 
 $s/S$ (bottom panel) and $\gamma_s$ (solid lines middle panel) 
as function of $\tau$, for   widely 
varying initial gluon conditions ($T$ in top panel, 
$\gamma_{\rm g}$ dashed middle panel), constrained to same entropy content.
}
\end{figure*}

We present our results for strangeness production in QGP
  in Figs.  \ref{TwoVol}--\ref{massdep}.
 In Figs. \ref{TwoVol} (RHIC) and   \ref{LHCVolVol2}  (LHC) we show the 
  centrality dependence, with the  two volume models,  
  see Eqs.\,(\ref{V1} and \ref{V2}),  corresponding to columns,  with V1 on left    
  and V2 (donut model)  on right. In the following 
  Figs.  \ref{Gluedep}-- \ref{Gluedep} we explore 
the dependence on the assumptions made. Here
  we show   RHIC results on left and LHC results on right.   
  In all Figs.  \ref{TwoVol}--\ref{massdep} we show three panels above each other. As noted already, 
  we show in the top panel, by solid line(s), the model time-temperature profiles. 
  The experimental observables are shown  as solid line(s) 
  in the middle panel  ($\gamma_s$) and in the bottom panel ($s/S$).
 The other lines illustrate as appropriate the key
inputs   used to obtain these results. When several lines of the same type
are present, we are presenting the impact parameter dependence, scaling 
the size and entropy content as discussed  above. 
In general, the temperature is followed down to a freeze-out at $T_f=0.14$ GeV.

In figure \ref{TwoVol}, we show what we believe is 
the  best $\gamma_s(\tau)$ (solid lines, middle panel) 
and $s/S(\tau)$ (solid lines, bottom panel) 
for RHIC  at 100+100 GeV at varying reaction centrality. The solid lines in the 
top panels show $T(\tau)$ for these 6 different centralities, with the lowest 
temperatures seen for the least central collisions, all temperatures
continue to $T=0.14$\,GeV. 
The slight increase in the initial temperature with increasing 
centrality is result of the 
scaling of initial entropy, which accommodates the observed change in 
$dS/dy\vert_{\rm f}$  beyond participant scaling, see Eq.\,(\ref{entsca}).
For all centralities (and below also for RHIC) we assume the same initial 
$s/S(\tau_0)=0.016$.
All lines shown begin at   $\tau_0=1/4$ fm, where the 
initial temperatures range $T_0\in (0.55 , 0.6)$ GeV. 
 For the V1 model, the  range of $\tau$ 
 spans the interval  $\tau_f=2.2$  fm (most peripheral) to
 $\tau_f=6.5$ fm (most central). In the donut 
  expansion model V2, this range is from  $\tau_f=3$ to 8 fm.

In the top panel, we also show the growth with $\tau$ of the transverse
expansion velocity (dotted line), and the strangeness 
pair energy threshold  $2m_s^r$ using running strange 
quark mass  (dashed lines). We note that the temperature 
drops below this threshold for the most peripheral reactions 
considered already at $\tau=1$ fm/$c$, and this occurs for the most
central reactions at $\tau=2 $fm/$c$, for model V1, and respectively, 
1.5 and 2.8 fm/$c$, for model V2. Thus, high strength thermal 
strangeness production life span  
varies by as much as factor 3, depending on centrality, and  
the expansion model. 

In the middle panel, we show dashed the rise of the gluon occupancy $\gamma_{\rm g}$ 
which we employed. The quark occupancy $\gamma_q$ is following the same 
functional temporal 
evolution  starting with 2/3 smaller initial value and evolving 
1.5 times slower. Because gluons dominate strangeness 
production in QGP, we do not show  $\gamma_q$ explicitly. 
The dotted lines show
how the volume evolves toward its maximum value at freeze-out. Each line is 
normalized to unity at freeze-out.  The actual value of the volume can be
read of the figure \ref{Volume}, given the value of $\tau$.

To summarize  the key results: we see a gradual increase of strangeness
yield with centrality, reaching near strangeness QGP equilibrium  for the
 most central collisions at RHIC. We have checked  stability of this result 
against variation of model assumptions. More detailed discussion
will be presented further below, see  e.g. Fig.~\ref{alfdep}.

\subsection{Strangeness production predictions for LHC}\label{LHC}
We performed a similar evaluation of strangeness production at LHC,
see figure~\ref{LHCVolVol2} which follows the same pattern as 
figure~\ref{TwoVol}. There are three modifications which 
were introduced when we consider LHC:\\ 
\indent (i) To account 
for the greater reaction energy, as already discussed, we 
increase the entropy $dS/dy$  by factor 4, which implies an assumed
increase in rapidity density of hadrons by a similar factor. We assume that
in elementary parton interactions the relative strength of strangeness
and non-strange hadron production is unchanged and thus, we  keep the 
initial relative yield $s/S=0.016$ constant.  Given the entropy yield
increase, we implicitly assumed  
  an increase in initial strangeness yield by a
factor 4 at LHC compared to RHIC.\\
\indent (ii)
In order to accommodate the greater transverse expansion pressure, 
we  increased the maximum transverse
flow velocity which can now attain $v_\bot=0.80$c (dotted line, 
top panel figure \ref{LHCVolVol2}).\\
\indent (iii)
We further assume that   thermalization time has dropped 
from $\tau_0=1/4$ fm at RHIC to  $\tau_0=1/10$ fm at LHC. However,
inspecting the slowly changing initial state evolution, in figure \ref{LHCVolVol2},
there would  be little  change in our results, were  $\tau_0$ to remain 
unchanged between RHIC and LHC. At this early time,  $\tau_0=1/10$, 
the value of $\gamma_s(\tau_0)$ at LHC is similar  
to the situation at RHIC, compare the beginning of the solid line in middle panel 
of  figures \ref{TwoVol} and \ref{LHCVolVol2}. 

This is so, since  the magnitude of the phase space scales with $T^3$ and the 
initial temperature $T(\tau_0)$ is considerably greater at LHC: 
in  the top panel, in figure \ref{LHCVolVol2}, we see that it
reaches up to $T=1.25$ GeV. For this reason, we have to show in the top panel
(dashed lines) $5 m_s^r$ rather than $2 m_s^r$, in order to fit it visibly into 
the top panel of the figure, and this is the only difference in 
the display of LHC results, in figure \ref{LHCVolVol2}, as 
compared to the RHIC results, figure \ref{TwoVol}.

We note that despite a much greater expansion velocity,  the evolution
time at LHC is significantly longer, with the most central collisions taking up
to 30\% longer to reach the freeze-out temperature, $T_f=0.14$ GeV.
The reader who prefers earlier freeze-out, at, {\it e.g.}, $T_f=0.17$ GeV can evaluate
the changes required by consulting the temperature profiles shown in the 
top panel.

The different centralities, at LHC, are considered with the same scaling 
of the transverse size and entropy content as we did for the case of RHIC. 
Comparing the left to right set of results ({\it i.e.}, volume V1 to donut V2 
expansion models)
for LHC, we see a more developed chemical equilibrium of strangeness with 
clear evidence of (over)saturation of yields for a few centralities, see the 
bottom panel on  right in figure  \ref{LHCVolVol2}.   There is  greater
final specific strangeness content at LHC than at RHIC, with visibly greater
thermal production leading to strangeness (over)saturation. 

At RHIC, the thermal production raises the 
value of $s/S$ from 0.016 to 0.028 for most central collisions
(V1 model of bulk volume expansion),
while at the LHC the thermal production raises $s/S$ from
 0.016 to 0.032.  We will discuss
below what this increase means for the $K/\pi$ and other particle ratios. 
The same relative increase
in $s/S$ is seen  in the model V2 of the expansion comparing  RHIC and LHC. However, if
the homogeneous bulk expansion applies at RHIC, but a donut type expansion
 arises at LHC,
the increases in strangeness yield and lifespan  would be more spectacular. Depending
on its expansion dynamics, LHC clearly harbors the potential to surprise us.

\subsection{Study of the dependence on initial thermalization condition}\label{initial}
An important question is how the value of  the
   unknown initial conditions impacts 
the results we  presented above. We have studied this question in depth 
in many different model approaches. The answer `practically no dependence' 
is best illustrated in the figure \ref{Gluedep}, where 
 we show the more conservative volume expansion model V1 results on left 
for RHIC, and and on right for LHC. We explore
 a wide range of initial gluon (and quark) occupancy $\gamma_{\rm g}$, 
which for consistency with other figures is shown in the middle panel 
by dashed lines, the initial values we consider for glue occupancy
vary  as $0.1<\gamma_{\rm g}(\tau_0)<2.1 $ in step of 0.5.  The second
of these lines, from the bottom, is the reference behavior we use in  
figures \ref{TwoVol} and \ref{LHCVolVol2}. 

We recall that, with 
$\gamma_{\rm g}(\tau_0)$, we also  vary $\gamma_q(\tau_0)$,
which  following the same  functional temporal 
evolution  starting with 2/3 smaller initial value and evolving 
 1.5 times slower.   Note also that the scale,
in top panel, varies between RHIC (left) and LHC (right) cases 
 the dashed lines denote  $2m_s$ on left, and   $5m_s$ on right. 
 Since the initial value of 
$s/S=0.016$ and $dS/dy= 5,000$ on left for RHIC  and, respectively 
$s/S=0.016$ and $dS/dy=20,000$ on right for LHC  is set, there is a 
corresponding variation in $T_0$ (top panel, left end of solid lines) 
and $\gamma_s$ (left end of solid lines in middle panel). The final 
results for   $\gamma_s(\tau_f)$ (right end of solid lines in middle panel)  
and $s(\tau_f)/S$ (bottom panel) are impressively insensitive
to this rather exorbitant diversity of initial conditions at fixed entropy content.  
The spread in $s/S(\tau)$ we see in the bottom panel could be seen as a wide 
line width.

\begin{figure*}[t]
\vskip -0.5cm
\psfig{width=7.9cm,figure= 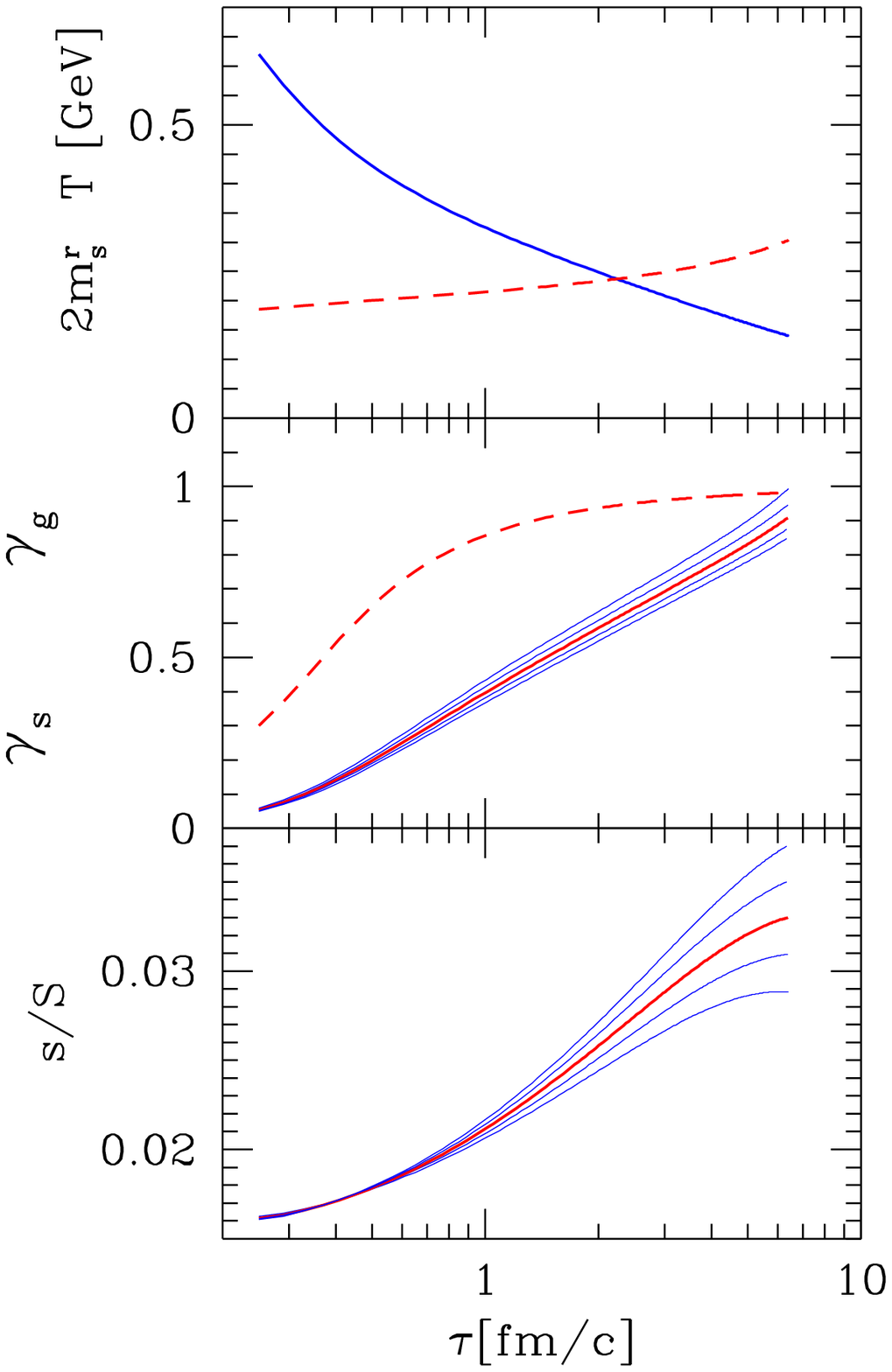       }
\psfig{width=7.9cm,figure= 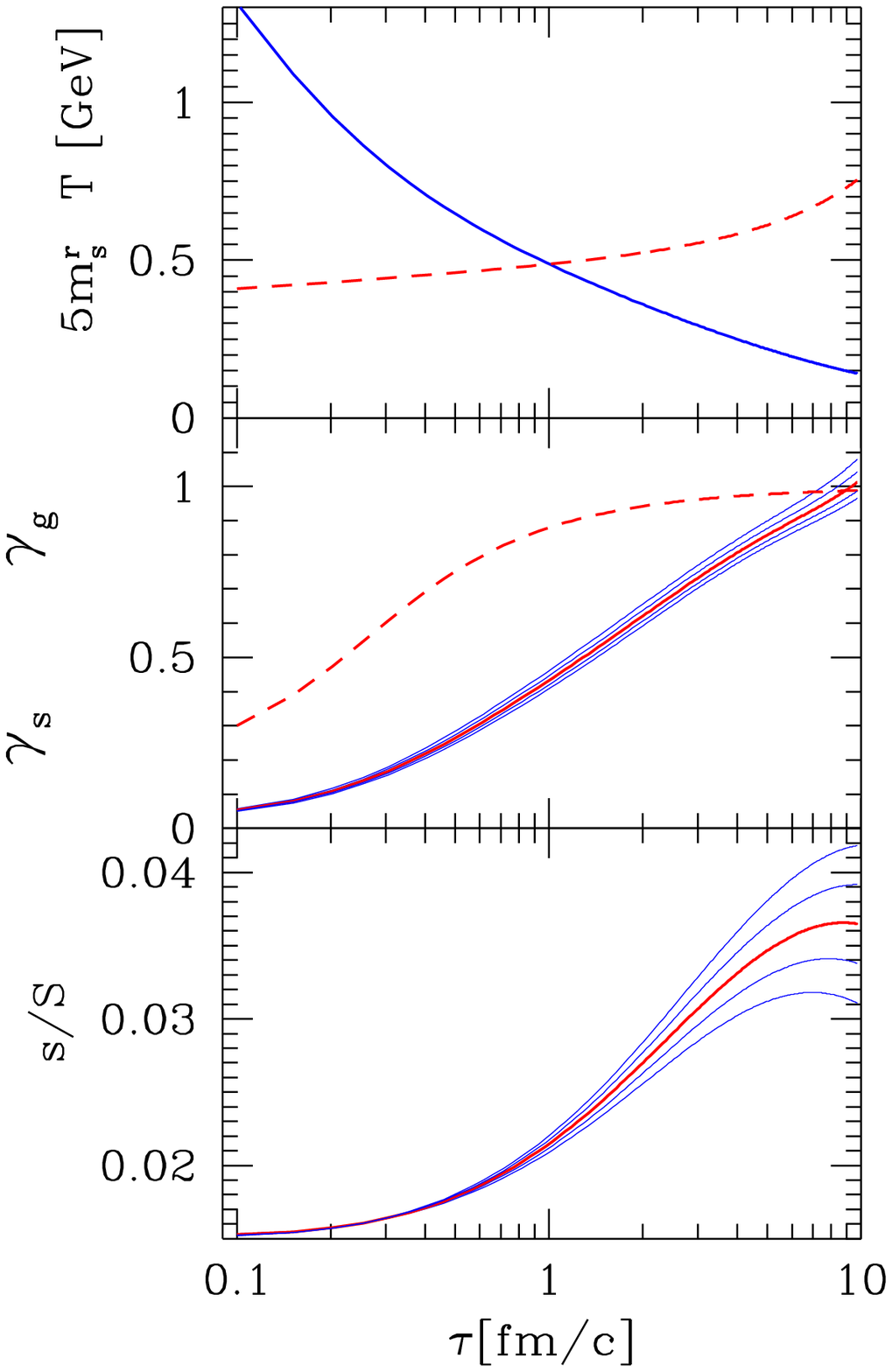   }
\vspace*{-0.6cm}
\caption{\label{alfdep}
(color online) Study of   of $s/S$ evolution with $\tau$,
on left for RHIC, on right for LHC, V1  volume expansion model.
Figure structure same as figure \ref{Gluedep}. 
Middle panel, solid lines: computed evolution of $\gamma_s$ in the deconfined phase
for values of $k=2,\,1.5,\,1,\,0.5$ and 0, see Eq.\,(\ref{gs});
Bottom panel: corresponding  evolution of $s/S$. The lowest  $\gamma_s$ line
corresponds to $k=0$ and the largest to $k=2$,   the opposite applies 
for $s/S$ lines. 
}
\end{figure*}
 
We conclude that strangeness cannot probe the very initial QGP conditions  near
$\tau_0$ ---  the memory of the initial history of the reaction is lost, the system
is opaque  for $\tau<2$--3fm/$c$ to the strangeness signature. On the other hand,
and most importantly, for the study here undertaken, 
this also means that 
experimental observables are characteristic of the properties of nearly
chemically equilibrated QGP. 

However, there remains dependence on the history of the QGP
fireball in models in which gluon (and light quark) chemical equilibrium
is not attained even in central reactions at 
RHIC~\cite{Pal:2001fz,He:2004df}. Thus only if 
gluons in  QGP did not approach the chemical equilibrium 
at $\tau_f$, a signature  of this condition would be seen in the strangeness
yield as is seen considering the right hand
 side of Eq.\,(\ref{qprod3a}). 
Correspondingly smaller value of $s/S$ are  then 
expected. 

Yet,  our here presented results agree well with current RHIC strange hadron production 
results regarding the total strangeness yield. This constitutes  indirect
evidence for the achievement of light quark and gluon chemical equilibrium 
in QGP formed in the most central, highest $A$  and highest energy  RHIC reactions.
 On the other hand, the relatively
small size and small lifespan of a  QGP potentially also 
formed in peripheral collisions could hinder the achievement
of QGP chemical equilibrium at hadronization.   
 
\begin{figure*}[t]
\vskip -0.5cm
\psfig{width=7.9cm,figure=   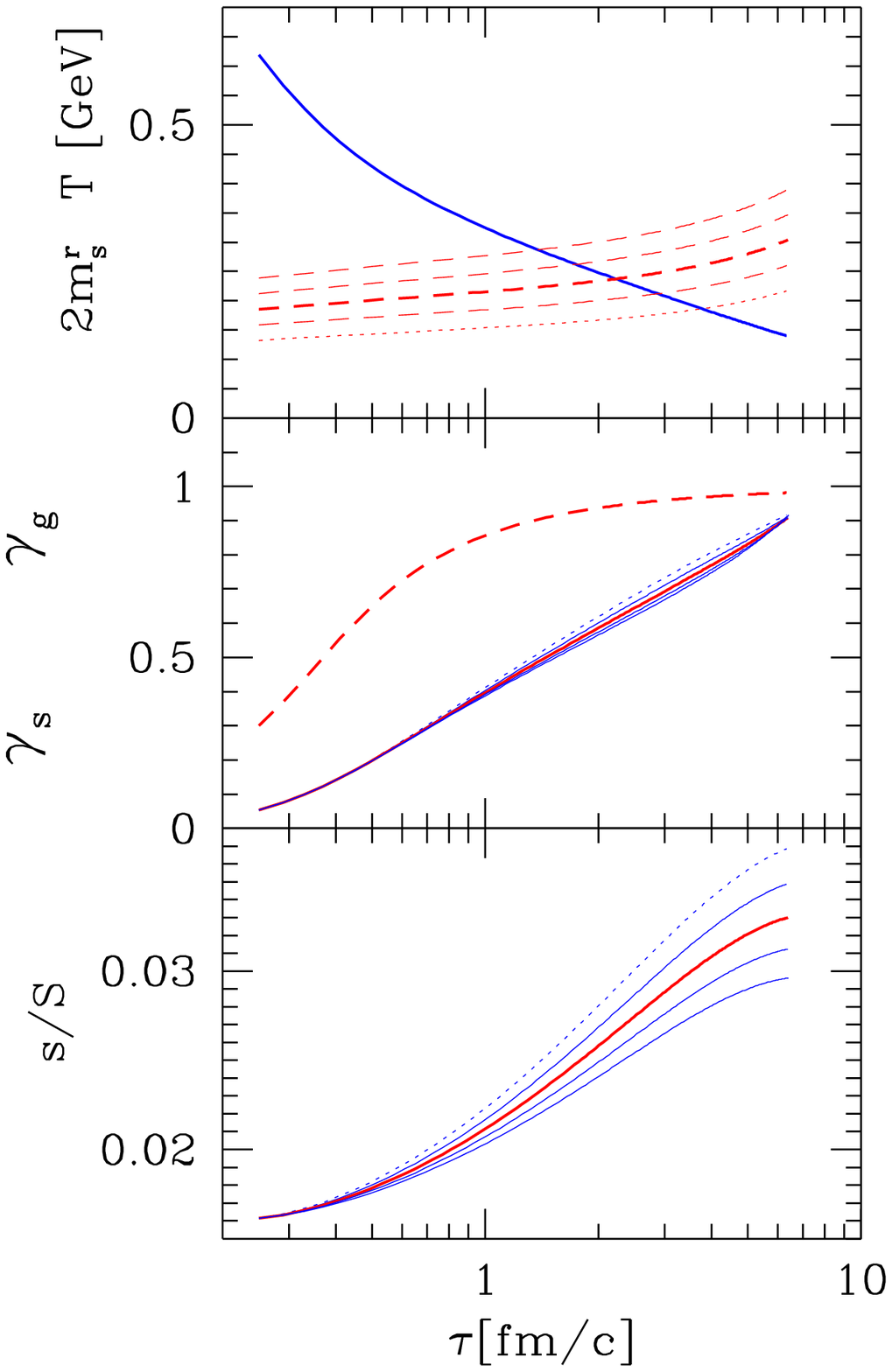 }
\psfig{width=7.9cm,figure=   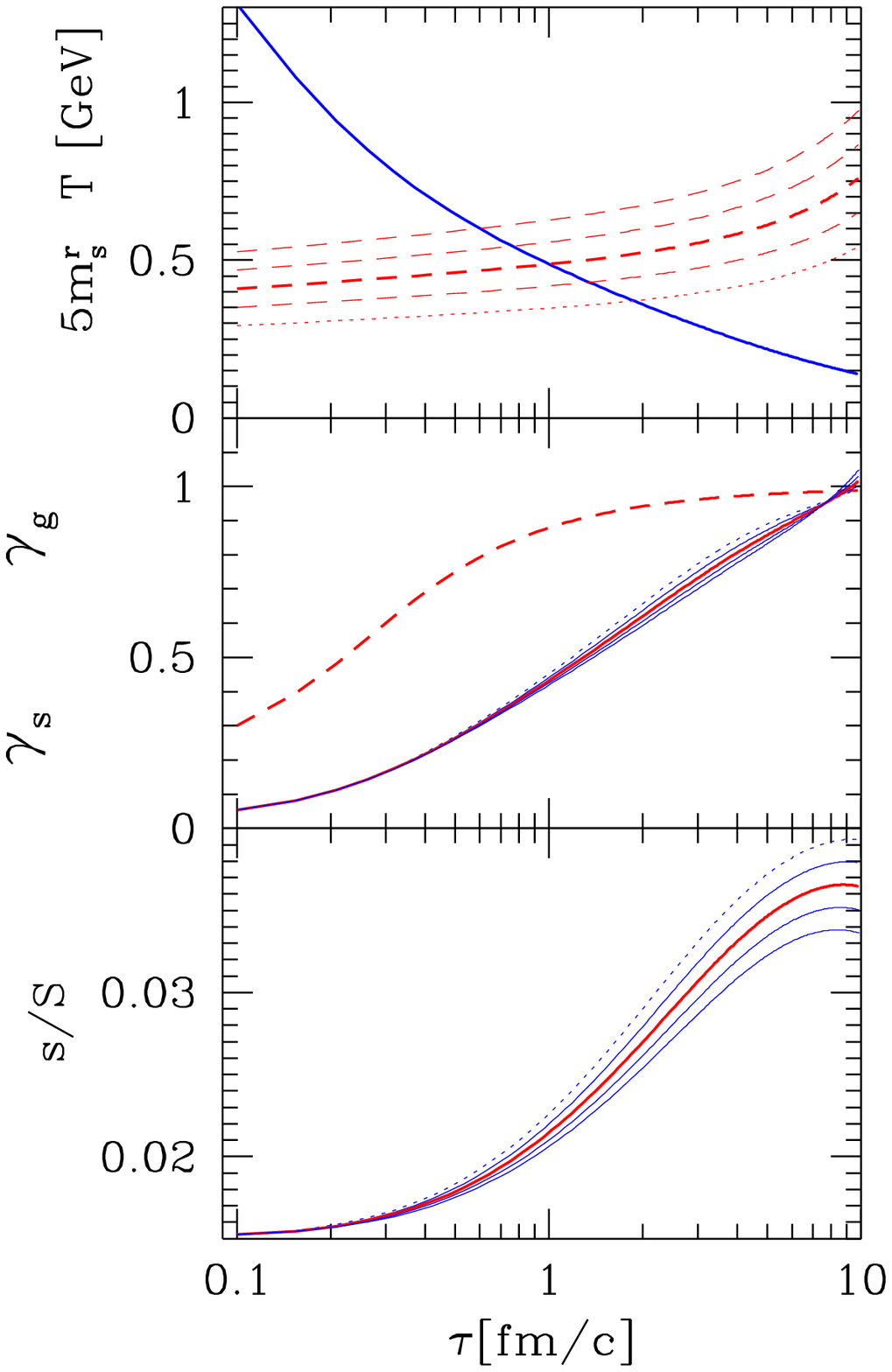  }
\vspace*{-0.6cm}
\caption{\label{massdep}
(color online) Study  of $s/S$ evolution with $\tau$ for different $m_s$,
on left for RHIC, on right for LHC, V1  volume expansion model.
Figure structure same as figure \ref{Gluedep} and \ref{alfdep}.
Top panel: 5 (running)  strange quark mass $m_s^r$. The middle of 5 lines
being our standard reference value, see text for more detail.
}
\end{figure*}

\subsection{Fundamental uncertainties in strangeness production}\label{Fund}
There are two QCD related uncertainties in our strangeness production 
study which we explore in turn in figures \ref{alfdep} and \ref{massdep}:\\
\indent 1. the effect of interaction
on the number of strange quark degrees of freedom, Eq.\,(\ref{gs}), \\
\indent 2. the value of strange quark mass. \\
We will now show that these uncertainties lead to observable effects,
in particular regarding the final value of $s/S$, and to a lesser
extend also regarding the   final 
value of   $\gamma_s$ (chemical yield equilibration). 

We first recall that 
$\gamma_s$ is introduced in Eq.\,(\ref{sdens}) 
in order   to relate the prevailing strangeness
density to the chemically equilibrated density at 
temperature $T$. Given a strangeness yield, the value of $\gamma_s$ depends on 
the properties of QCD by the way of what the chemically equilibrated density is.
The   actual value of $\gamma_s$  enters decisively
into the kinetic equation of strangeness production. Eq.\,(\ref{qprod3a}) shows
that the smaller the  value of $\gamma_s$, 
the bigger is the obtained change in the value of $s/S$.

Since, in our approach, we used Boltzmann
statistics for the strangeness degree of freedom, the effect
of Pauli blocking on strangeness production is not considered and in
fact is a minor effect.
The direct dependence  of physical observables on $\gamma_s$ arises 
from the process of strangeness reannihilation into gluons. We note that 
Eq.\,(\ref{qprod3a}), describing the change in strange quark yield,  can also 
be written in  the form:
\begin{equation}\label{qprod3aex}
\frac d{d\tau} \frac s S = a_{\rm g}
(\rho_{\rm g}^2\rho_s^{\infty\,2}\,-\rho_s^2\rho_{\rm g}^{\infty\,2} )
+a_q
(\rho_q^2\rho_s^{\infty\,2}\,-\rho_s^2\rho_q^{\infty\,2} ),
\end{equation} 
where instead of $\gamma_i$  the actual densities of particles appear. We see that 
for each particle both the equilibrium, and transient, density 
must enter in order for the system  to attain chemical equilibrium as function
of time. Consequently, the value of $\gamma_s$ and hence the  QCD correction
in Eq.\,(\ref{sdens}) matters. 

The figure \ref{alfdep}, which follows the pattern of
figure \ref{Gluedep}, illustrates this effect when the effective
degeneracy, see Eq.\,(\ref{gs}), varies. 
We consider for values of $k=2,\,1.5,\,1.,\,0.5$ and 0. $k=2$ is
the perturbative effect seen for massless quarks, $k=0$ corresponds
to no effect of interaction, when $m_s\gg T$. 
When $k=0$, the strange quark degeneracy is largest, thus 
for a given strangeness yield $s$ the value of $\gamma_s$ is smallest.
and the production of strangeness biggest. Consequently, this
value corresponds to the smallest $\gamma_s$ in the middle panels of 
figure  \ref{alfdep} and greatest value of $s/S$ in the bottom panels. 
Other lines follow, and the middle solid lines (red on line ) correspond to 
$k=1$, which we used to obtain the reference results presented earlier.

A similar effect arises considering variation of strange quark mass as
shown by dashed lines in top panels of figure \ref{massdep}.  
We vary by factor 2 the strange quark mass
as is seen in the top panel of the figure \ref{massdep}, on left for 
RHIC and on right for LHC. 
When the strange quark
mass is increased, the equilibrium strangeness density is decreased, 
and thus, for a given  strangeness yield $s/S$, the value of $\gamma_s$ is
increased, which in turn reduces strangeness 
production strength. Smallest mass considered
(dotted line in top panels of figure \ref{massdep}) corresponds to the 
largest final value of $s/S$ shown by dotted line in the bottom panels. 
What is more surprising is that the effect of mass variation {\it cancels}
in the actual computed $\gamma_s$ shown in the middle panels. 
The reason for this accidental cancellation 
is that for a larger mass the smaller value of final $s/S$ solves for the same
value of $\gamma_s$, see Eq.\,(\ref{sS1}).
 
Since both $m_s$, and the interaction effect on the strange quark effective
degeneracy, see Eq.\,(\ref{gs}), are today not understood at sufficiently precise level, 
the appearance of a possible  range
of values at freeze-out, for both $\gamma_s$ and $s/S$, 
in figures \ref{alfdep} and  \ref{massdep}, signals (correlated) uncertainty
in the understanding of the results at RHIC and predictive power for LHC.

\section{Consequences for hadron   yields and their evolution from RHIC to LHC}\label{eval}
\subsection{Strangeness and entropy}\label{sh}
The final value of $s/S$  is the key result which practically alone
determines  relative strange
particle yields. It depends on the value of $dS/dy$ which we 
start with. The value we take,  $dS/dy=20,000$  at LHC (4 times RHIC), is a 
guess arising from extrapolation of the energy dependence of 
particle production. However, it fixes in effect the initial 
conditions, and leads to the range of values for  $s/S$ we obtained. 
We had found that there is continued growth of strangeness yield beyond
the RHIC energy range where $dS/dy=5,000$, which suggests a further strong increase
in the strange hadron yield we will address below.

\begin{figure}[t]
\psfig{width=7.9cm,figure=  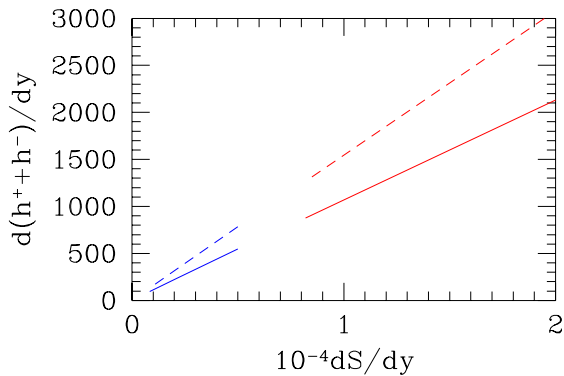  }
\vspace*{-0.6cm}
\caption{\label{hsS}
(color online) The yield of charged hadrons $d(h^-+h^+)/dy$ 
for different values of $dS/dy$, left domain for RHIC and
right upper domain for LHC.
Solid lines: before weak decays, dashed lines: after all weak decays. 
}
\end{figure}

Variation of $dS/dy$ amounts to variation of the 
produced hadron yield.  We expect that $dS/dy\propto dh/dy$.
To quantify this, we present, in figure \ref{hsS}, how the yield of
charged hadrons relates to the entropy yield. These results were
obtained both for RHIC (left region, red on line) and LHC (right region)
by the  methods described in  Ref.~\cite{Rafelski:2005jc},
employing  SHARE  suite of programs~\cite{Torrieri:2004zz}. 
In this evaluation 
we have  available  for each $s/S$ (at RHIC and LHC) the corresponding 
strangeness occupancy  $\gamma_s$. We further can fix the 
ratio of charge to baryon number as in the incoming nuclei, $q/b=0.39$.

A further difference
between RHIC and LHC is that  we take the thermal energy per baryon at RHIC
to be $E/b=39.3$ GeV and following Ref.~\cite{Rafelski:2005jc} 
$E/b=412$ GeV at LHC. The consistent
ranges for $s/S$ at RHIC are $0.018< s/S< 0.03$ and, at LHC, we can address
$0.018<s/S<0.037$, we cannot otherwise find a smooth match of QGP to HG-phase space
within the physical range of phase space occupancies $\gamma_s,\,\gamma_q$.  
We note that this LHC choice, $E/b=412$ GeV,
limits the range of possible entropy yield to just below the range
we explore in our present work,  $dS/dy=20,000$, four times the RHIC range.

\begin{figure*}[t] 
\vskip -0.5cm
\psfig{width=7.9cm,figure=  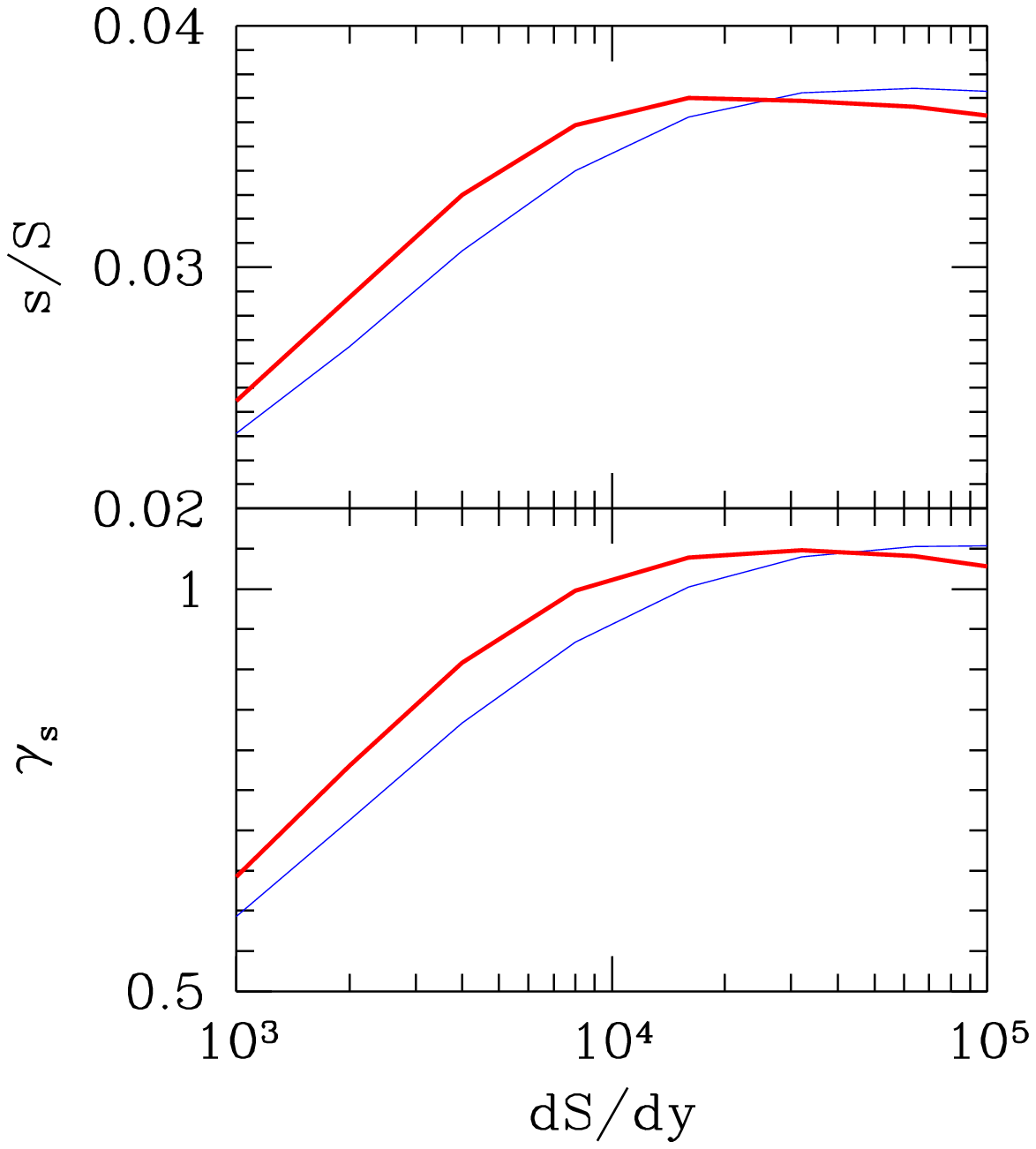   }
\psfig{width=7.9cm,figure=   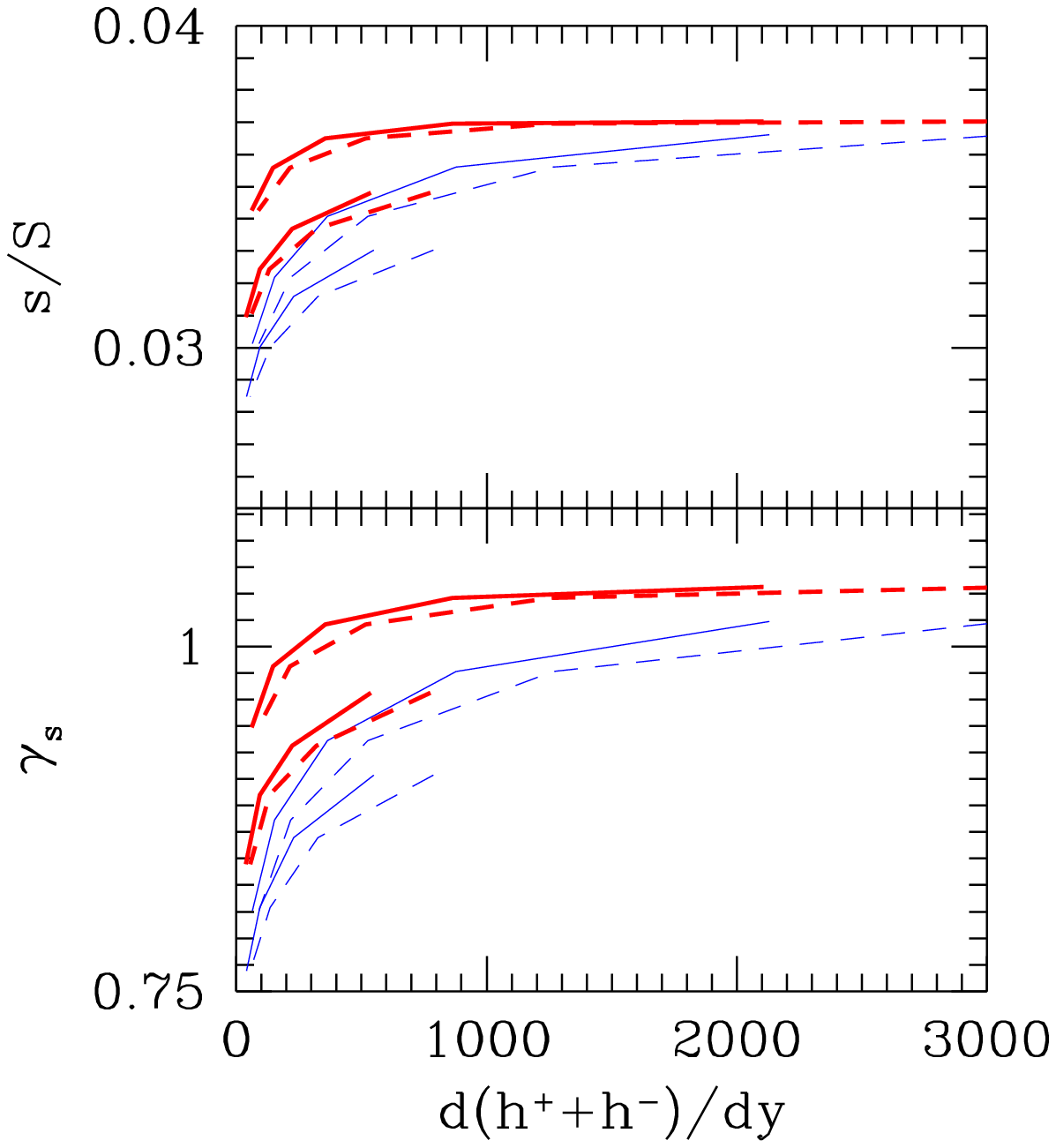   }
\vspace*{-0.6cm}
\caption{\label{sSdvdydhdy}
(color online)   $s/S$ (top panel) and $\gamma_s$ (bottom panel) 
for the most central 5\% collisions
as function of   $dS/dy$ (on left) and as function of  $d(h^++h^-)/dy$
on right.
The bulk transverse expansion model V1 is the thin solid line (blue), 
the thick line (red) is 
the donut expansion model V2 with $d=3.5$ fm. 
On right, solid lines: before weak decays, dashed lines: after  weak decays
(excluding K$^\pm$ see text). 
We recognize the RHIC domain results by the smaller $dh/dy$ 
range of results presented.
}
\end{figure*}

We next explore the resulting final $s/S$ and
$\gamma_s$.  We show these quantities, on left in figure \ref{sSdvdydhdy},   
as function of the $dS/dy$ input,  and
as function of the  charged hadron multiplicity $d(h^-+h^+)/dy$,
on right. The two expansion models we consider are as before 
V1 homogeneous expansion  (thin solid, lower line) and 
V2 donut expansion model (thick solid, upper line).
We see a gradual rise of both $s/S$ and $\gamma_s$ 
as function of $dS/dy$ which begins to saturate  for
  $dS/dy>20,000$\,, but at a rather high values beyond 
chemical equilibrium: the expansion is so fast that there
is no time to reannihilate the very abundant 
strangeness before freeze-out.

In  Fig~\ref{sSdvdydhdy} on the right hand side, we see 
the RHIC (smaller $dh/dy$ range)  and LHC  domains. 
Dashed lines are the charged hadron multiplicity after weak
decays, in particular of neutral strange hadrons such 
as K$_{\rm S}$. The negligible impact   of charged 
Kaon weak decays to hadron  multiplicity is not accounted
for as this effect is small and experiment dependent considering 
the quasi-stability of K$^\pm$. 
In the RHIC domain, we have a nearly linear rise of $s/S$ (top
panel) and  $\gamma_s$ (bottom panel) with $ d(h^++h^-)/dy$.
In the LHC domain chemical (over) saturation 
of these quantities is clearly visible.
The   left, and right hand side of  figure \ref{sSdvdydhdy} 
have at large $dS/dy$ and  $ d(h^++h^-)/dy$, respectively, 
a similar appearance indicating chemical (over)saturation 
with the increasing hadron yield. We recall that while the 
functional form of the figure is correct, the normalization of
the abscissa is somewhat arbitrary for the case  of LHC, which
is based on an extrapolation of SPS and RHIC results.

We next study the growth of strange hadron
yield   with $s/S$. In order to not confuse this
with the issue of final hadronization volume of QGP, we 
will study particle ratios. The method of evaluating 
particle yields is, as just discussed in study of the 
charged hadron yield: SHARE program is asked to match micro-canonical 
conditions in the QGP phase to those of HG phase space. 
Given the statistical parameters, we can compute all particle 
yields within the statistical hadronization approach. 

\begin{figure}[t]
\vskip -0.3cm
\psfig{width=7.7cm,figure=  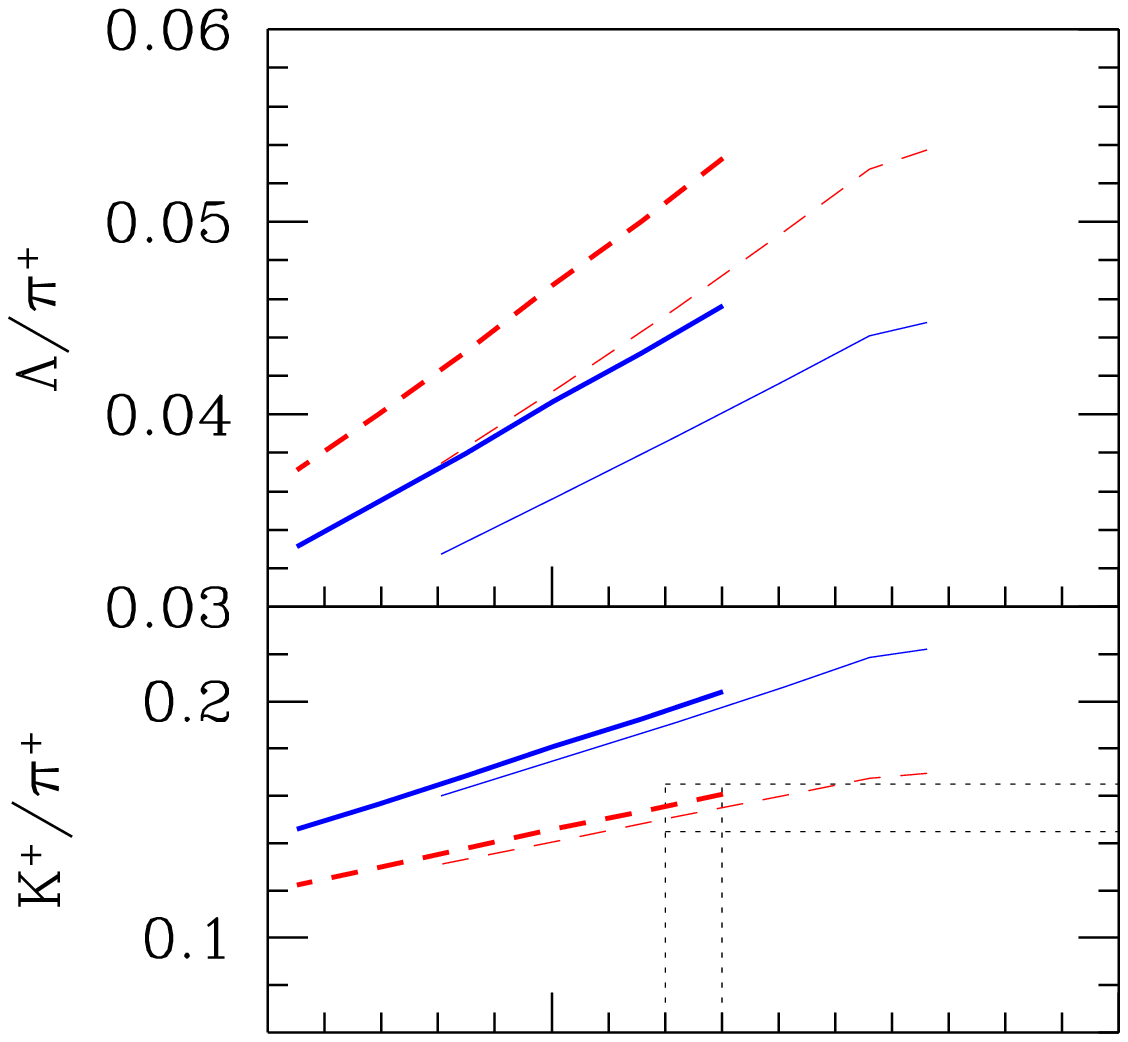   }\\
\vspace*{-1.7cm} 
\psfig{width=7.7cm,figure=  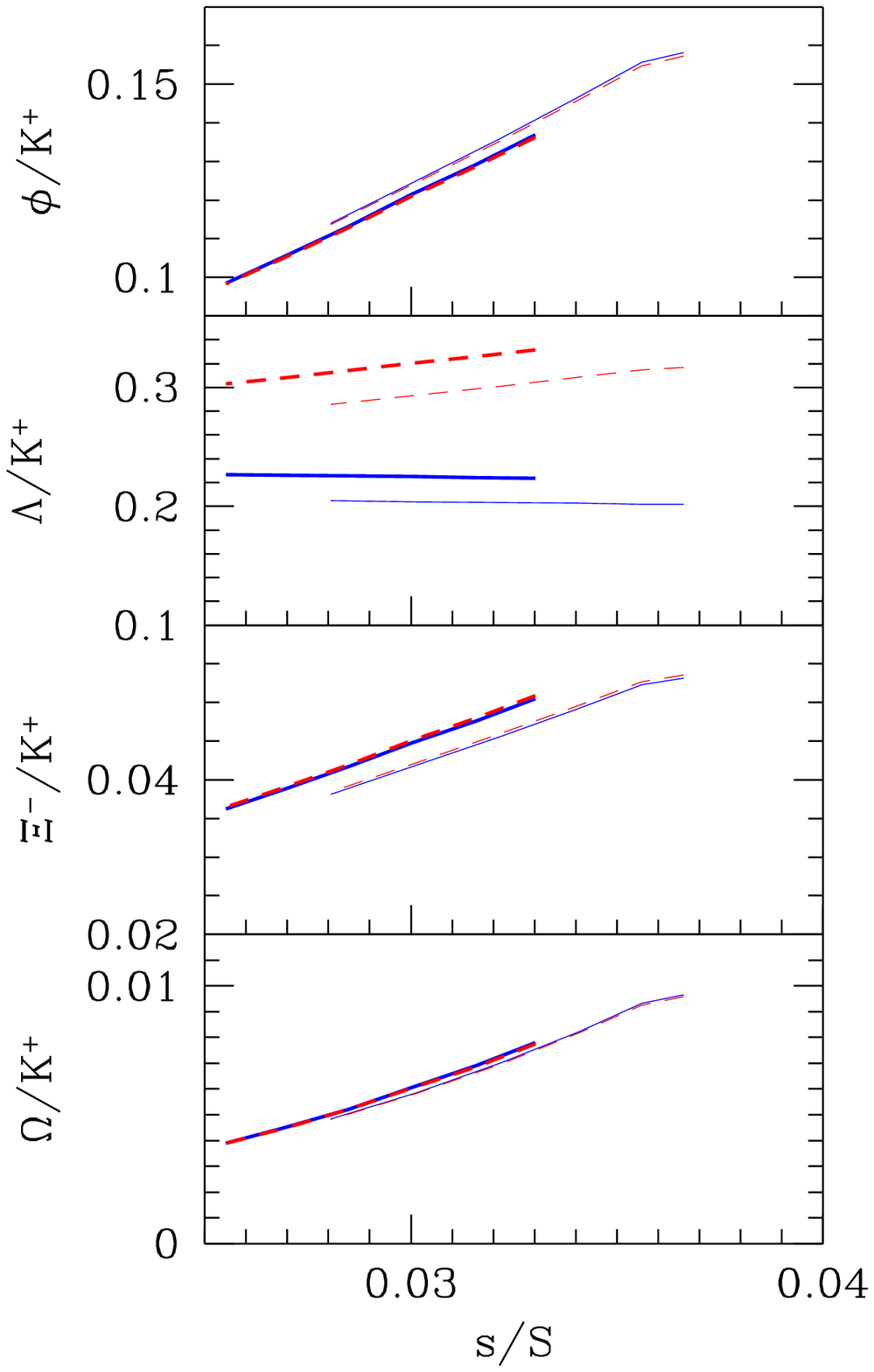  } 
\vspace*{-0.6cm}
\caption{\label{LKpFLXOsS}
(color online) Relative particle yields  as function of $s/S$:
from top to bottom $\Lambda/\pi^+$, K$^+/\pi^+$, and
$\phi/\mathrm{K}^+,\ \Lambda/\mathrm{K}^+,\ 
 \Xi^-/\mathrm{K}^+,\ \Omega^-/\mathrm{K}^+$.
 Solid lines (blue) are the primary relative yields, 
dashed lines (red) give the yields after weak decays (K$^+$ is not decayed). 
Thick line with $s/S<0.03$ are for RHIC, and thin lines with $s/S<0.037$
are for LHC physics environment. 
Dotted lines guide the eye for the RHIC ${\rm K}^+/\pi^+$ ratio.
}
\end{figure}

We first consider, in the two top panels of 
figure \ref{LKpFLXOsS}, the ratios $\Lambda/pi^+$  
and K$^+/\pi^+$   as function of $s/S$, the
attained specific strangeness per entropy at QGP  freeze-out. 
 Solid lines are before weak decays, and dashed 
lines present the corresponding results 
after weak decays. These enhance the pion
yield, but as we see, even more the $\Lambda$ yield. The K$^+$ yield is 
practically not  changed: considering the long life span of the charged
kaon it is common to present   experimental results corrected for
any decay. Thus, K$^\pm$ are considered, in our study, 
as if they were stable particles. The RHIC results are the thicker lines
and the LHC results are the thinner lines.

The RHIC and LHC results 
for K$^+/\pi^+$, $\phi/\mathrm{K}^+$ and $\Omega^-/\mathrm{K}^+$
are practically overlapping, since the influence of the difference in
baryochemical potential is not material for this ratio. However, the 
greater final $s/S$ yield expected in most central LHC collisions
implies an increased yield at LHC compared to RHIC. The dotted lines 
in the second from top panel, in figure \ref{LKpFLXOsS},  guide the eyes
 both in the RHIC domain, and at the higher $s/S$, LHC domain. 

We find, in this
perhaps easiest  to measure K$^+/\pi^+$ ratio, 
a noticeable and measurable relative 
yield increase. This prediction is important, since
after dropping with energy at SPS, this ratio remained of the same constant
magnitude (within error) at the much higher RHIC energy domain. 
There is  less change expected 
between RHIC and LHC for the $\Lambda/pi^+$ ratio, since the decreasing 
baryochemical potential is compensating, to a large extend, the effect of 
the increase in  strangeness yield. We note that there will be the  
opposite  effect in the $\overline\Lambda/\pi^+$ yield. 
 We see that the ratios $\phi/\mathrm{K}^+$ and $\Omega/\mathrm{K}^+$
are most sensitive to the change in $s/S$ occurring between RHIC and LHC.
The change in baryochemical potential diminishes this sensitivity considerably in the 
$\Xi/\mathrm{K}^+$ ratio.  Except for $\Lambda/\mathrm{K}^+$
the weak decays have a negligible impact on all ratios with $\mathrm{K}^+$ 
considered here.
 
\subsection{Strange particle yields as function of centrality}\label{PY}
We have obtained the growth of strangeness yield in QGP with centrality
and this allows us to explore participant dependence of strange 
hadron yields. Since we do not know well how $E/b$ and $E/TS$ changes
as centrality changes, we choose  to follow a different approach in
order to present our results. We assume that the hadronization 
occurs at a fixed condition, and first we consider $T=140$ MeV and 
$\gamma_q=1.6$, corresponding to supercooled sudden hadronization. 
In a second step, we compare these results to those obtained with 
$T=160$ MeV and $\gamma_q=1$, which is the equilibrium hadron phase space. 
The fixed value of statistical hadronization parameters as function of 
centrality follows the pattern seen in the analysis of 
RHIC results~\cite{Rafelski:2004dp}.

To obtain the results,  we follow a similar  computational 
scheme as developed  in the study of $s/S$ dependence, except that we 
now fix  $T$ MeV and  $\gamma_q$, and not, {\it e.g.}, $E/TS$. These 
conditions lead to similar hadronization conditions, but $E/TS$ had some variability 
for small participant number, see figure 4 in Ref.~\cite{Rafelski:2004dp}.

\begin{figure*}[t]
\vskip -0.5cm
\psfig{width=7.9cm,figure=  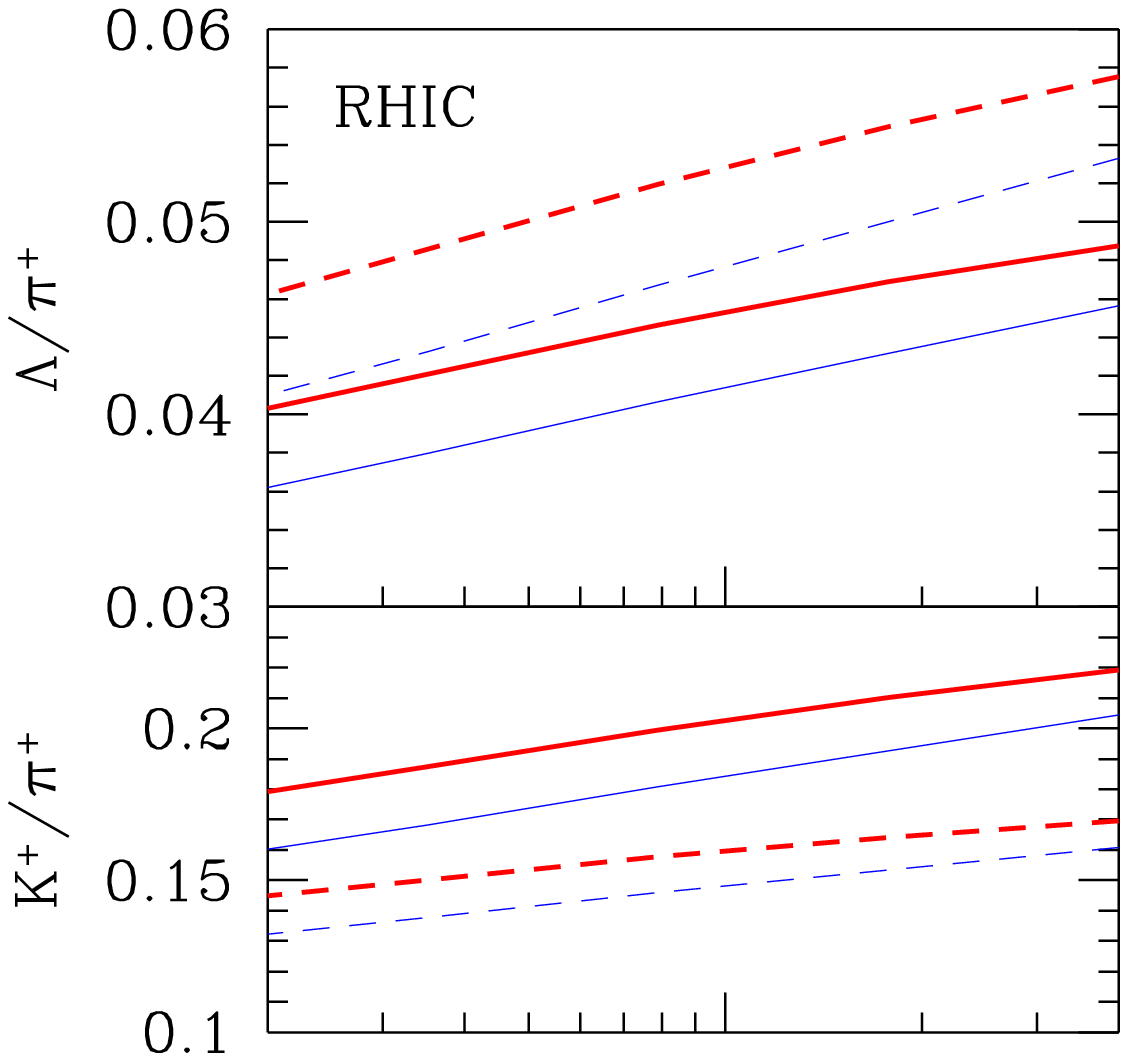    } 
\psfig{width=7.9cm,figure=  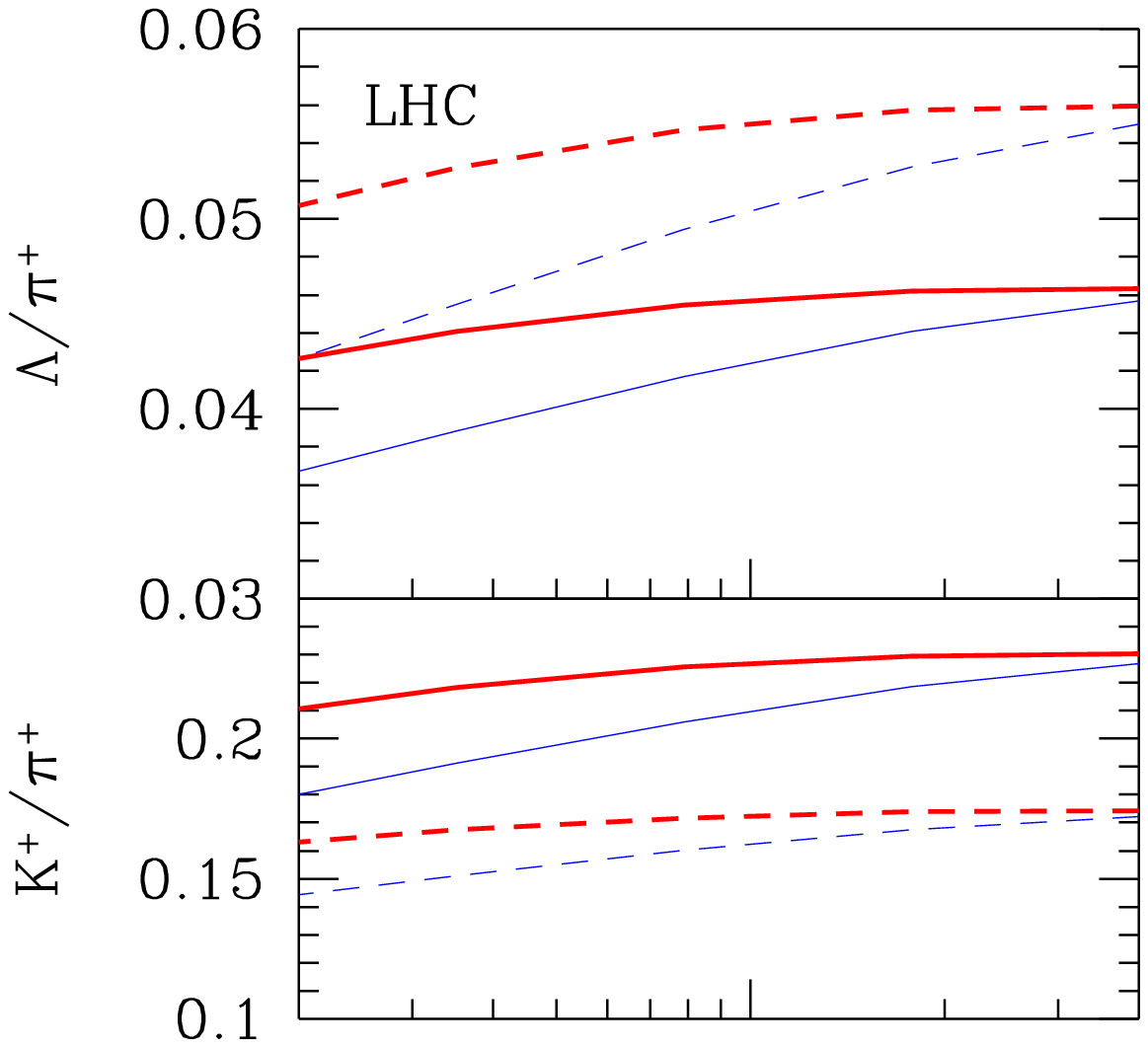   }\\
\vspace*{-1.3cm}
\psfig{width=7.9cm,figure=  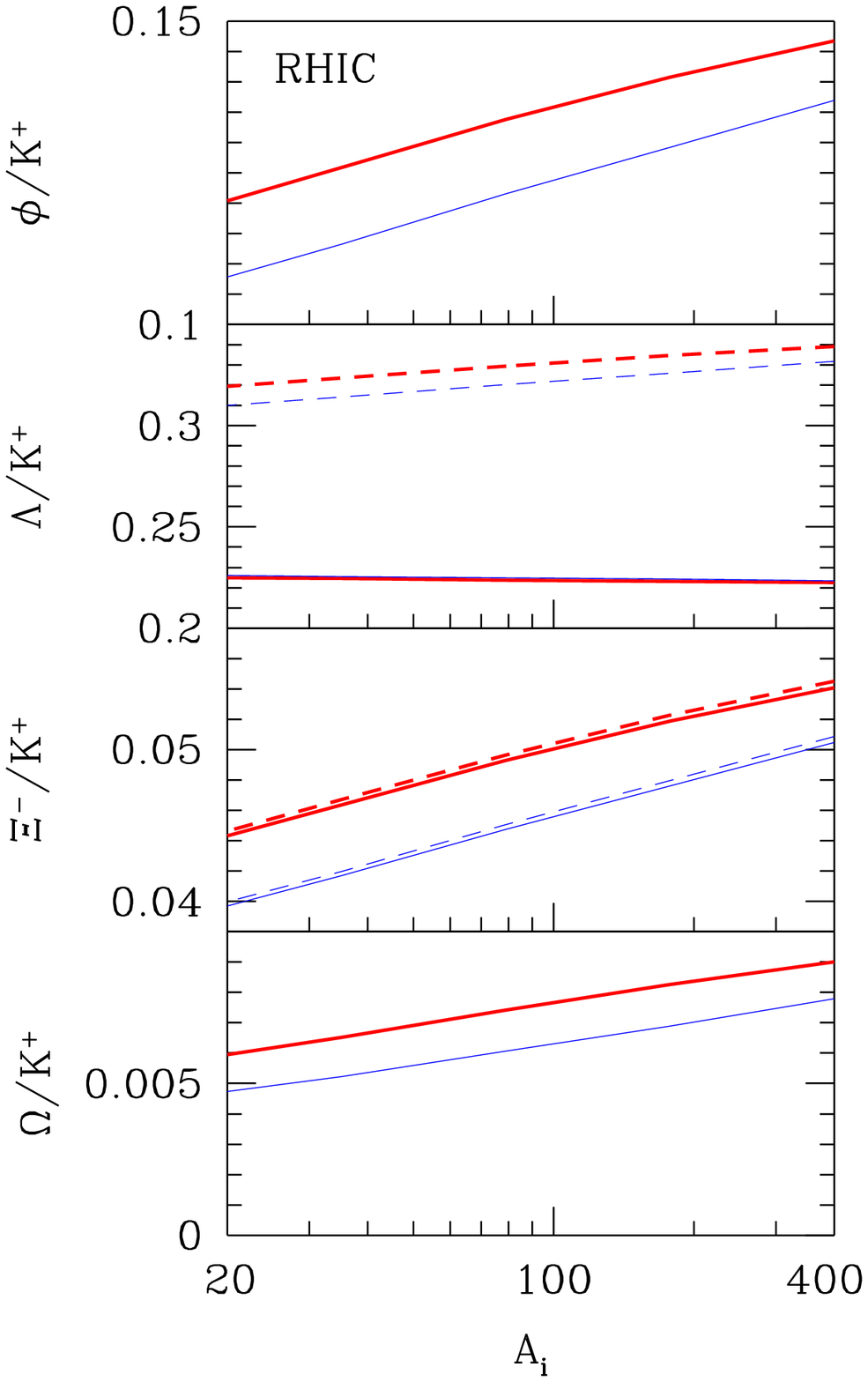   }
\psfig{width=7.9cm,figure=  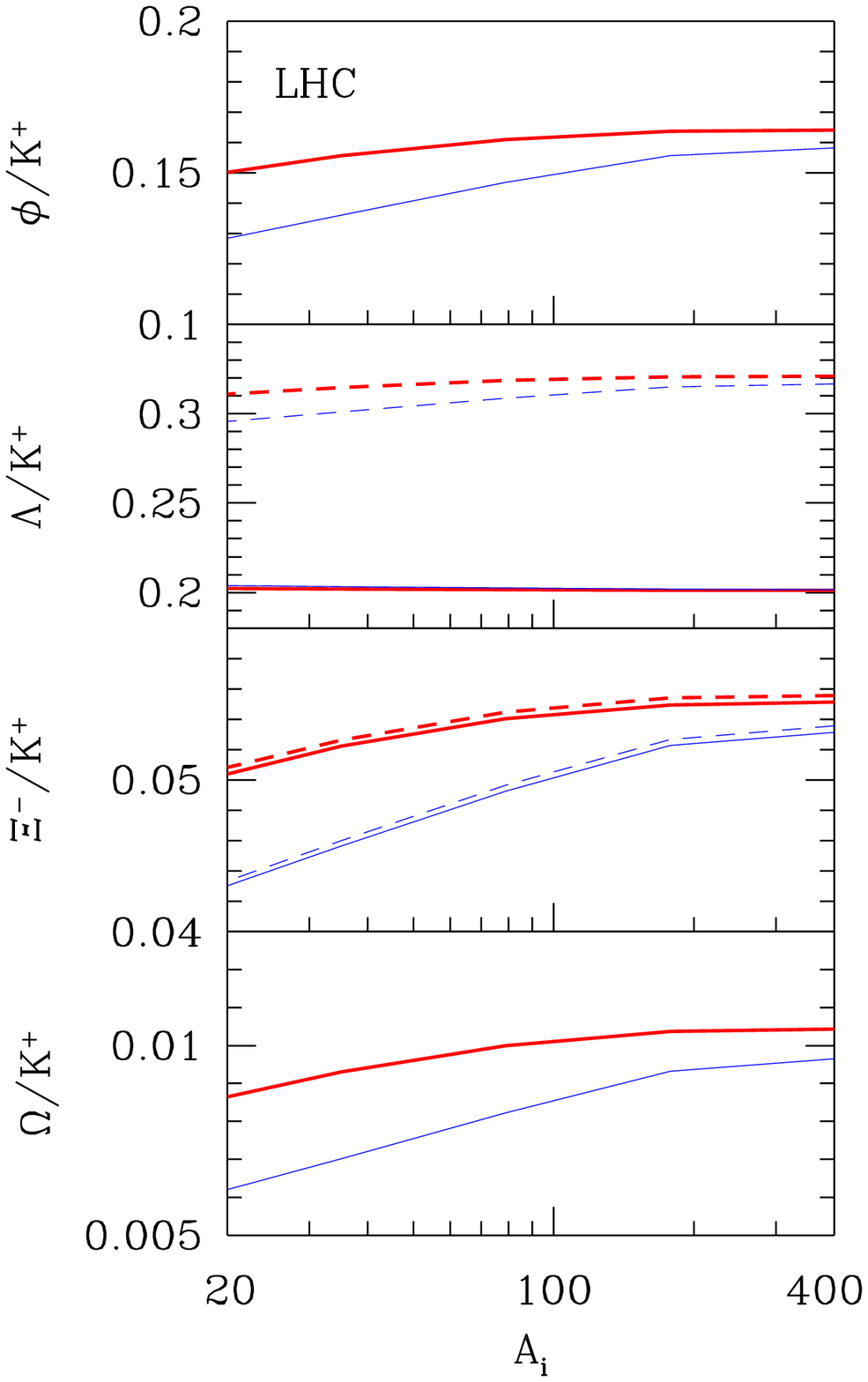    }
\vspace*{-0.8cm}
\caption{\label{LKpiA}
(color online)   Relative strange particle yields as function of 
participant number $A$, left RHIC and right LHC. 
 From top to bottom,  $\Lambda/\pi^+$, K$^+/\pi^+$, and 
$\phi/\mathrm{K}^+,\ \Lambda/\mathrm{K}^+,\ 
 \Xi^-/\mathrm{K}^+,\ \Omega^-/\mathrm{K}^+$.
 Solid lines primary relative yields, dashed lines (relative) yields
 after all weak decays (not shown when difference is within line width).  
Thin lines (blue), model V1 (volume 
expansion) and thick lines (red), model V2 (donut expansion).  
Results are for   supercooled sudden hadronization.
}
\end{figure*}

In the figure \ref{LKpiA}, we show the results,
on left for RHIC, and on right  LHC. We follow a similar  display
scheme as in figure \ref{LKpFLXOsS}, and 
show  $\Lambda/pi^+$, K$^+/\pi^+$, followed by
$\phi/\mathrm{K}^+,\ \Lambda/\mathrm{K}^+,\ 
 \Xi^-/\mathrm{K}^+,\ \Omega^-/\mathrm{K}^+$,
as function of participant number. The thinner (blue) lines are 
for the V1 model of volume expansion, and the thicker (red) lines for the V2 model
of donut expansion. Dashed lines include the weak decays, which increase the 
$\Lambda/pi^+$ ratio and decrease the K$^+/\pi^+$ ratio. 

We see, in figure \ref{LKpiA}, a steady, but rather slow, increase with centrality
of the relative strange hadron, to non strange pion yield. 
The decay pions tend to flatten the  K$^+/\pi^+$ ratio at RHIC. We obtain
predict more rise of $\phi/\mathrm{K}^+$ than is observed at RHIC~\cite{Adams:2004ux}, but
our variation is  within the error bar of the experimental, nearly constant result, 
$\phi/\mathrm{K}^+=0.15\pm0.03$.  
As discussed earlier,   the overall increase in $s/S$ 
expected at LHC (right side) compared to RHIC (left side) explains
 the noticeably greater relative K$^+/\pi^+$ ratios at LHC. 
The most noticeable rise, with centrality, is expected when the ratio
has the largest disparity in strangeness content.  There is no 
centrality dependence expected when there is no difference in 
strangeness content, such as in $\Lambda/\mathrm{K}^+$, where
the weak decay produces very small centrality dependence. 
 
\begin{figure*}[t]
\vskip -1.5cm
\psfig{width=7.9cm,figure=  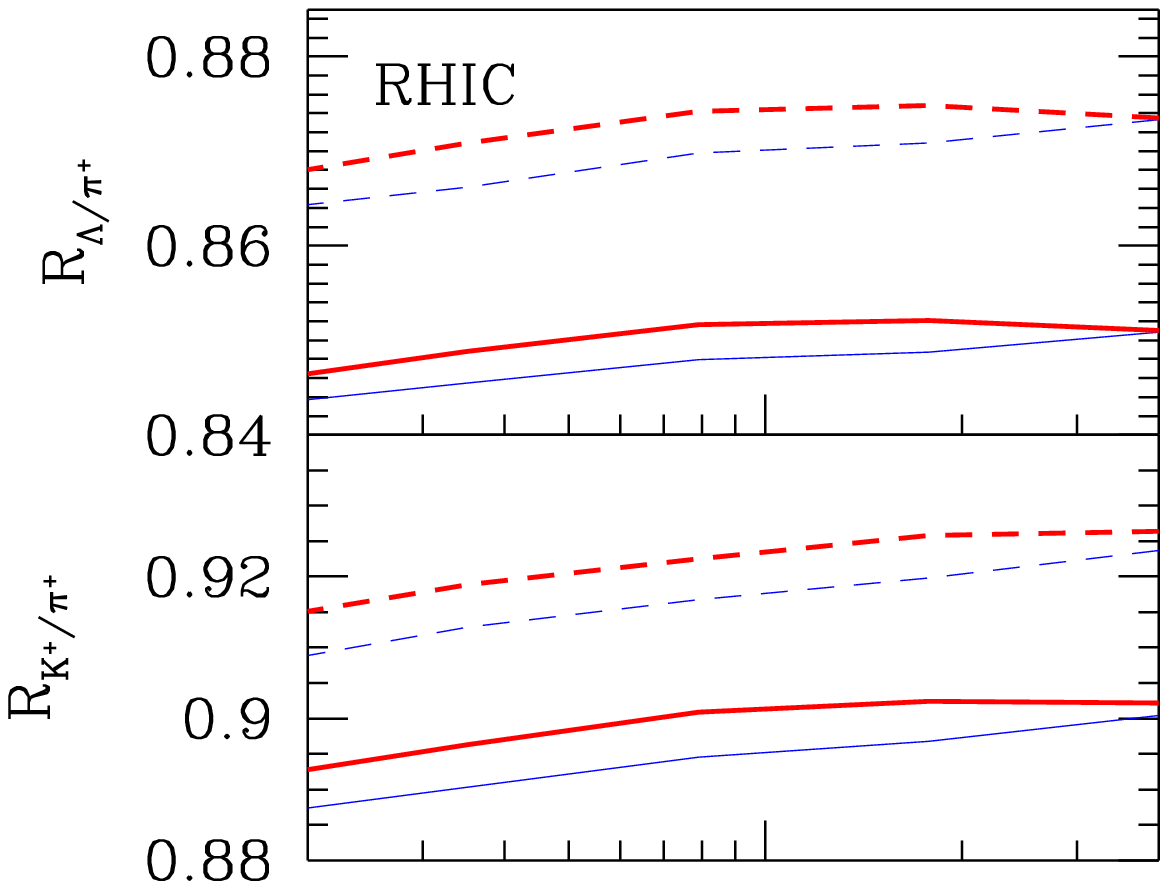     } 
\psfig{width=7.9cm,figure=  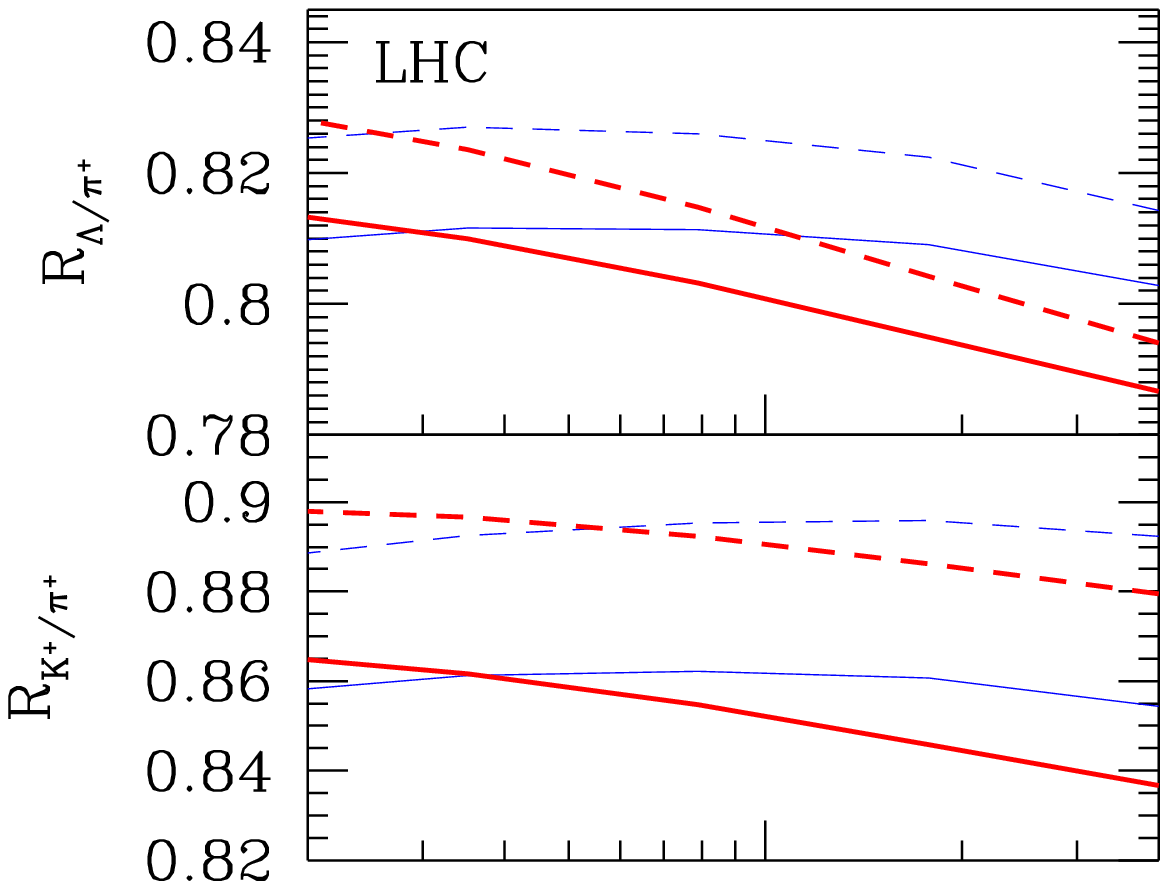    }\\
\vspace*{-1.7cm}
\psfig{width=7.9cm,figure=  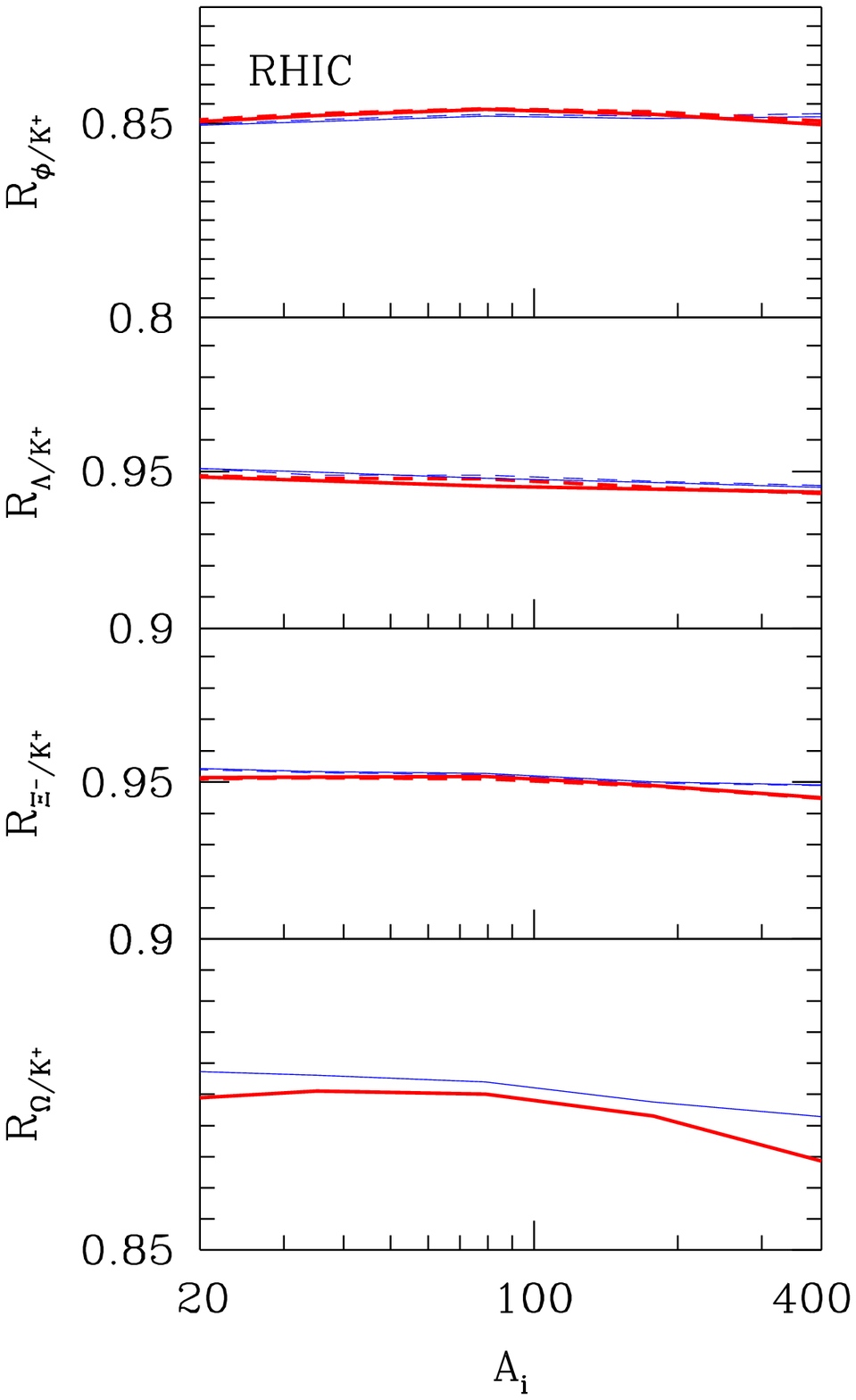    }
\psfig{width=7.9cm,figure=  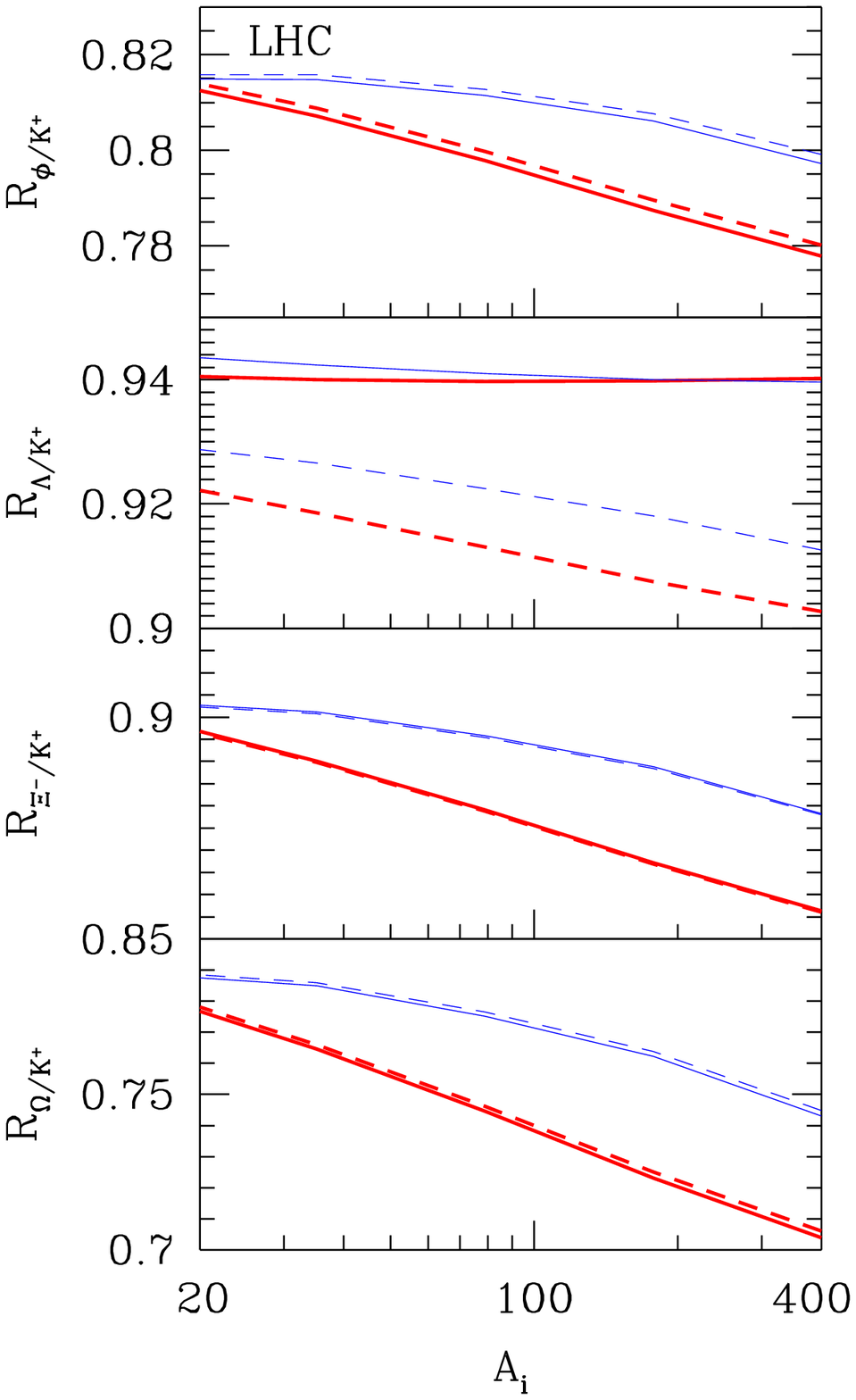     }
\vspace*{-0.6cm}
\caption{\label{LKpiART}
(color online)  Ratio of relative yields obtained  for sudden and equilibrium
hadronization (see text) as function of centrality. Left RHIC, and right LHC. 
 From top to bottom,  $\Lambda/\pi^+$, K$^+/\pi^+$ and
$\phi/\mathrm{K}^+,\ \Lambda/\mathrm{K}^+,\ 
 \Xi^-/\mathrm{K}^+,\ \Omega^-/\mathrm{K}^+$.
 Solid lines primary relative yields, dashed lines (relative) yields
 after all weak decays (when absent, no difference within line width with
solid lines).  Thin lines (blue), model V1 (volume 
expansion) and thick lines (red), model V2 (donut expansion).  
}
\end{figure*}

One of the interesting questions is how sensitive is the study of these particle
ratios to the hadronization conditions. In figure \ref{LKpiART}, we compare the 
sudden hadronization at $T=140$ MeV and $\gamma_q=1.6$ with the hadron 
phase space equilibrium model at  $T=160$ MeV and $\gamma_q=1$ forming
the ratio of particle yield ratios, that is the  results obtained for 
sudden hadronization are divided by  those obtained 
for the equilibrium HG phase space. The panels follow the same particle
ratios as in figure \ref{LKpiA},  on left we show RHIC and on right LHC
results. 

The value of $\gamma_s^{\rm H}(T=160,\gamma_q^{\rm H}=1)$ varies 
as function of centrality $1.26<\gamma_s^{\rm H}<0.88$. In general, the 
higher hadronization temperature assumed in this  chemical
equilibrium case favors the yield of the
more massive hadron. Thus,  ratios  of heavy to lighter particle
evaluated at the same value of $s/S$ as shown in  figure \ref{LKpiA} are 
bigger for the $\gamma_q^{\rm H}=1$ since we took a higher $T=160$ MeV 
chemical equilibrium freeze-out value.

We note that  the nonequilibrium model
is much better explaining available multistrange hadron data. 
The equilibrium model needs to be
amended to explain the  enhanced $\Omega$ yield. The way out
from this dilemma if one insists on HG equilibrium could be  
a   multi-freeze-out  temperature interpretation. The freeze-out 
temperature of the $\Omega^-$ in the equilibrium freeze-out 
model would need to be noticeably higher than that of the 
bulk of  strange particles. We note that such multi-freeze-out models
could experience systematic difficulties as function of centrality.   
Moreover,  considering  figure \ref{LKpiART}, in the chemical equilibrium
interpretation of hadron production a separate freeze-out for both $\Omega^-$ and
$\Xi^-$ will be required at LHC. Aside of being unpalatable, 
such a multi-freeze-out HG equilibrium model 
will not describe fluctuations well~\cite{Torrieri:2005va}.

\subsection{Thermal charm at RHIC and LHC}\label{thc}

\begin{figure*}
\vskip -0.5cm
\psfig{width=7.7cm,figure=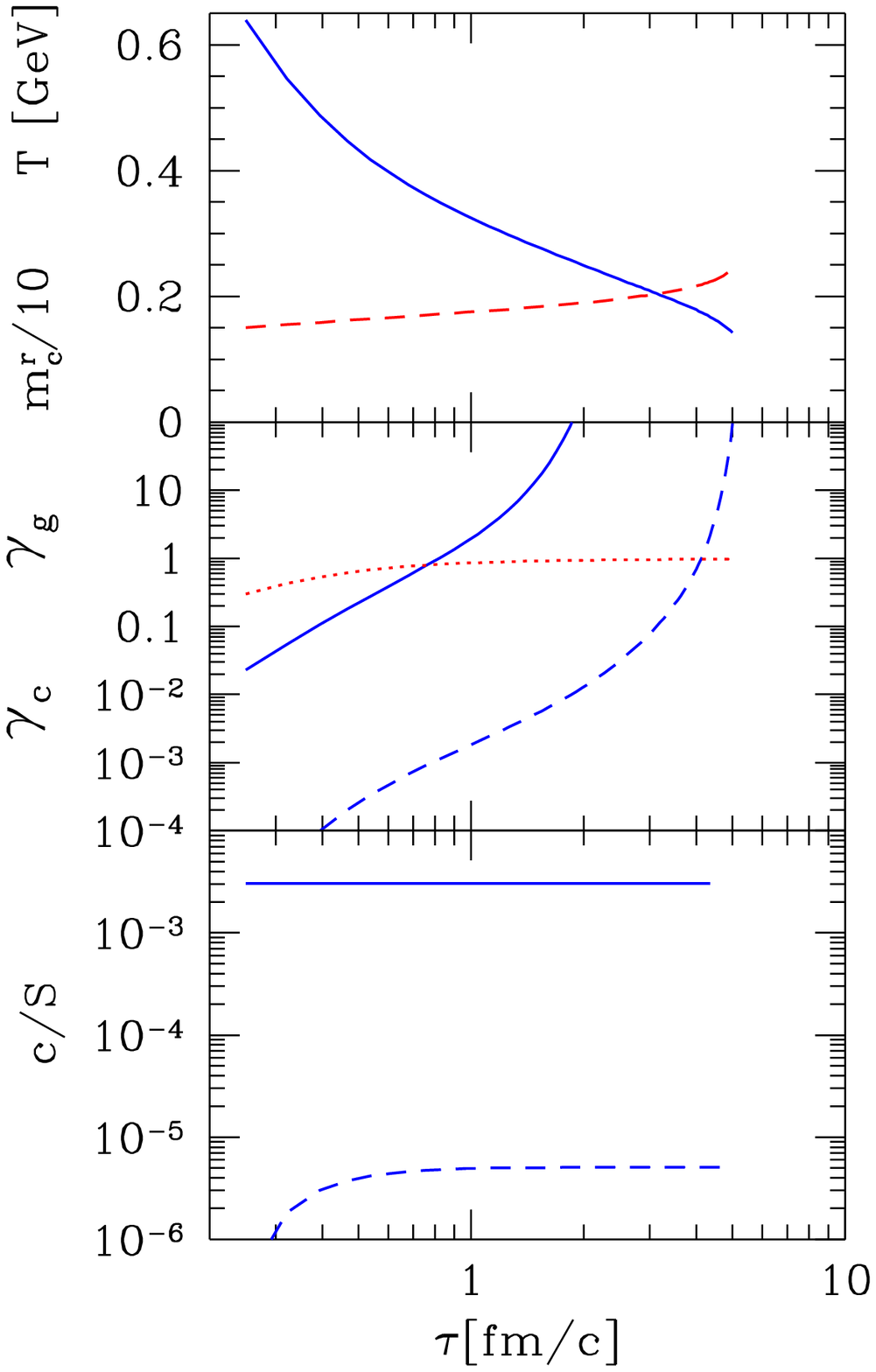 }
\psfig{width=7.7cm,figure=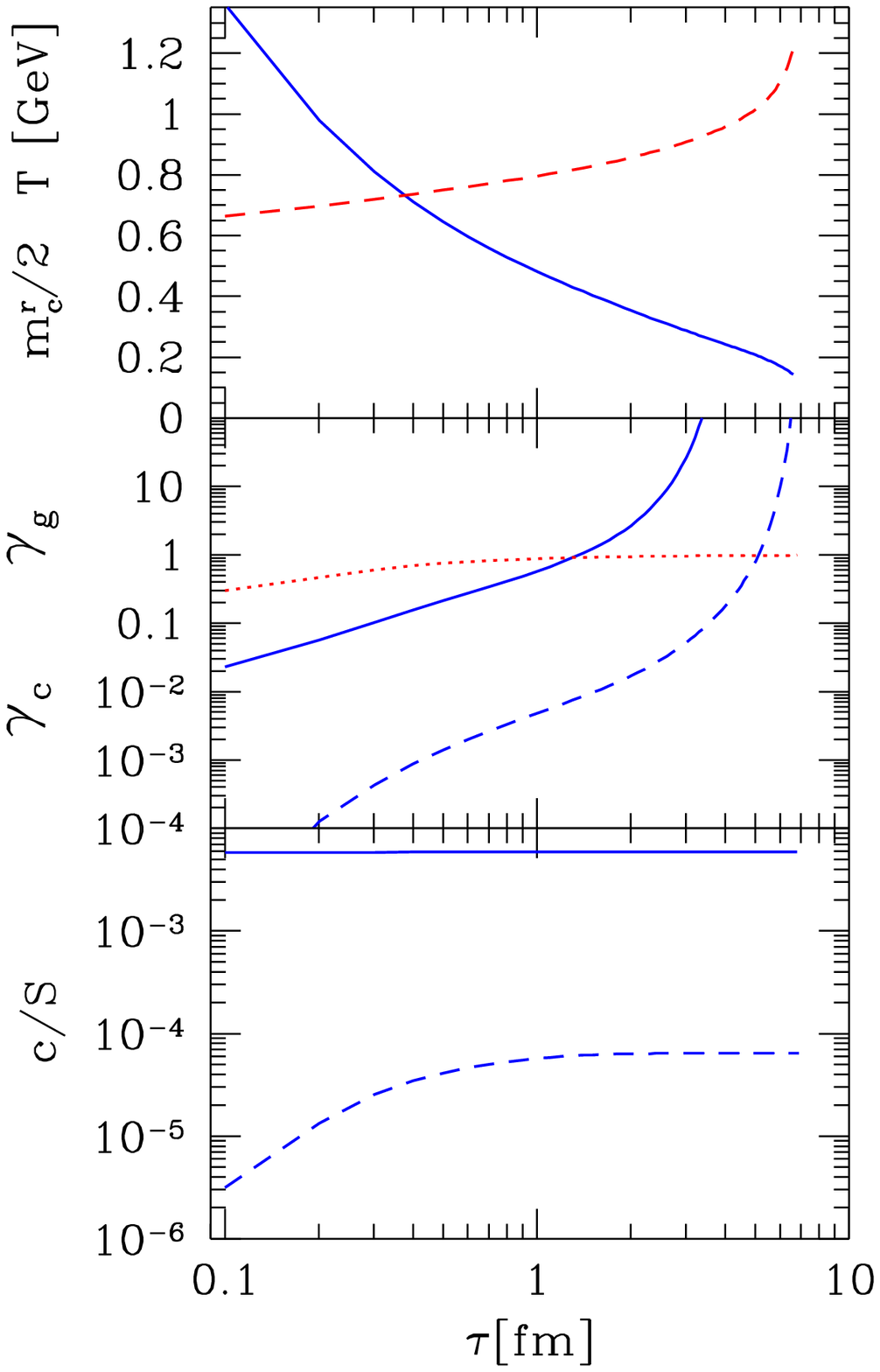 }
\vspace*{-0.5cm}
\caption{\label{Charm2Vol}
(color online) Left RHIC and right LHC for charm production. 
Figure structure same as figures \ref{TwoVol}--\ref{massdep}. 
Top panel:  solid lines  $T$, dashed lines, running $m_c^r$, scaled with 10  for
RHIC on left, and with 2 on right for LHC.
Middle panel: dotted line $\gamma_{\rm g}$, solid  lines the computed total charm 
$\gamma_c$, dashed lines $\gamma_c$ corresponding to thermal charm production.
Bottom panel: specific charm yield per entropy, solid lines for 
all charm, and dashed lines for thermally produced charm. 
 }
\end{figure*}

The direct  initial high energy parton 
collisions dominate production of the  massive flavor, charm and bottom,  
and a thermal process seems to be of no interest. However, 
there are two questions we can investigate:\\
a) considering that $\gamma_c$ in the deconfined phase can be as large
as $\gamma_c\simeq 100$ is there any significant thermal annihilation of charm in
the QGP evolution?\\
b) how large is the thermally produced charm yield and can it lead to chemical
equilibrium of charmed quarks? \\
The interest in question a) needs no further discussion. Question b) is of interest since
the directly produced charmed quarks are in principle not easily thermalized,
while the thermally produced  charmed quarks emerge naturally in a
 momentum distribution,
imaging their `parent' particles thermal momentum distribution. 
 Consequently, the thermally produced charmed quarks provide 
a solid thermal lower limit for the yield of charm, with the 
directly produced charm contributing to thermal distribution 
after charm has been subject to  
collisions required for thermalization. 

We see the results of this study 
in figure \ref{Charm2Vol}, on left for RHIC, on right for LHC.
The top panels, as usual, presents the temperature  and, 
charmed quark mass, scaled with factor 1/10 at RHIC and 1/2 at LHC (on right).
The middle panel presents  $\gamma_{\rm g}$,  the charm 
phase space occupancy, $\gamma_c$ and  $\gamma_c$ obtained solely
by {\it thermal processes}. Similarly, in the bottom panels, the dashed lines
are the thermal yield at RHIC and LHC, while the horizontal lines are 
the (little) evolving $c/S$ yields including the directly produced charm. 
The direct charm production,  at RHIC, is expected to be 
600 times greater than the thermal process. At LHC, the higher initial 
temperature, but unchanged  specific direct yield $c/S$, in parton collisions, 
suggests that  thermal production is just factor  90 times smaller. However, 
this factor depends on good understanding of both processes and the initial
conditions and surely cannot be fully trusted. Moreover, there is 
the possibility that the directly produced charm at LHC may not well thermalized,
in which case an appreciable fraction of thermal charm yield could indeed
originate in thermal reactions. 

The answers to the opening questions thus are\\
i) There is no visible charm reannihilation, 
see the nearly horizontal lines at $c/S=3\cdot 10^{-3}$ (left, RHIC) 
and $c/S=6\cdot 10^{-3}$ (right, LHC), in the bottom panel; 
note that the direct charm yield, we assumed implicitly, is 
obtained by multiplying the $c/S$ yields 
with $dS/dy=5,000$ on left for RHIC, yielding $dc/dy|_{\rm RHIC}=15$,
 and with $dS/dy=20,000$ on  right for LHC, yielding $dc/dy|_{\rm LHC}=120$.\\ 
ii) In the middle panel, we 
see that  thermal production alone (dashed lines) oversaturates the charm phase space,
the thermal produced charm phase space occupancy $\gamma_c^{\rm th}$
(dashed lines middle panels) cross  the gluon dotted 
$\gamma_{\rm g}$ line  at around 4 fm/$c$ for RHIC 
(corresponding to $T\simeq 0.175$ GeV) and at  
5 fm/$c$ (corresponding to $T\simeq 0.20$ GeV) at LHC.

\section{Summary and conclusions}

We have studied  the thermal QGP based 
  strangeness production  at RHIC and LHC,
and have interpreted the  observed final $s$-yield in terms of   
our theoretical knowledge about the properties of the QGP phase. Our aim 
has been to understand how the overall final state strange quark flavor 
has been produced, and to study in detail the mechanisms behind
strangeness enhancement. As a further objective we have
explored the impact of high strangeness on the strange hadron yields. 

Our results suggest that strangeness enhancement 
could be studied considering:
\begin{equation}\label{RsCP}
R^s_{\rm CP}\equiv {s/S|_{\rm central} \over s/S|_{\rm peripheral} }
      = {s/S(\tau_f) \over s/S(\tau_0) }.
\end{equation}
The central strangeness yields is just the final value we find 
at freeze-out, combining the initial yield with the additional 
thermal production. The peripheral yield is the initial value 
before thermal strangeness production begins. Our study shows that 
$R^s_{\rm CP}\in [1.6, 2.2]$, with the precise result depending
on details such as strange quark mass, see figure~\ref{massdep},
reaction energy and dynamics of expansion, see figure~\ref{Volume}.  

More generally, to separately consider $s$ and $S$ 
 we can use $d(h^++h^-)/dy$ as a measure of entropy $dS/dy$ 
content, see  figure~\ref{hsS}. Instead of the total strangeness, 
on may consider enhancement of individual (multi)strange particles, 
which we discussed in depth both as function of the achieved
$s/S$, in subsection \ref{sh},
and as function of centrality $A$ at fixed hadronization condition
for all centralities, in subsection \ref{PY}.

We note that the overall growth of the enhancement of the  
 strangeness yield with  centrality, at the level  
a  factor 1.6--2.2 is accompanied by a further   enhancement 
of  multistrange hadron yields, as is seen comparing the yields of 
multistrange hadrons with the yield of kaons. One can show that
the $\phi/\mathrm{K}^+$ ratio is mainly dependent on value of 
$s/S$ and not on hadronization temperature
(see appendix B2i ~\cite{Kuznetsova:2006bh}) 
when the strangeness conservation constraint is
implemented. This effect is unique to $\phi$ and
 arises since $m_\phi\simeq 2m_{\rm K}$. 

Aside of centrality dependence, we have  explored, within the 
framework of our model, the extrapolation 
from RHIC to LHC physics environments. More generally, 
we   have presented, in figure \ref{sSdvdydhdy}, the reaction energy dependence 
by considering the  central rapidity $ s/S$
yields as function of  $dS/dy$.
  
One of the interesting results obtained is the approach to chemical 
strangeness equilibrium in the deconfined QGP phase formed in most central
and high energetic RHIC reactions. The evidence for this 
is implicit in the experimentally reported yields of strange hadrons, which 
lead to values of specific strangeness per entropy at the 
level of $s/S\simeq 0.028$~\cite{Rafelski:2004dp}. Our study of QGP 
based kinetic strangeness production
provide an explanation of this result, both, the value of $s/S$ 
and as function of centrality.

Our present   study further shows
 that  the proximity of chemical strangeness
yield equilibration in QGP formed at RHIC and LHC, and 
the  effective opacity of  QGP to this signature, is the reason that considerably less
sophisticated models of QGP evolution which we, and others, have considered
are equally successful in the study the strangeness 
production, as long as these models yield 
conditions near to chemically equilibrated QGP.

Given the near chemical equilibration at RHIC, and within the models considered,
we obtain some strangeness over saturation at LHC. Moreover,  for the most central 
5\% reactions there is no relevant dependence of strangeness
production  on initial conditions prevailing in the reaction. We have
demonstrated this  in a picture-book fashion, see figure \ref{Gluedep},
where, for a wide range of initial conditions,  the same 
final strangeness yield and equilibrium condition arises after $\Delta\tau=$2--3 fm/c.

On the other hand, the more peripheral reactions do 
not saturate the phase space, in that both $\gamma_s^{\rm QGP}<1$, 
and $s/S<0.03$. Thus, the  peripheral yields, being sensitive to 
the initial conditions, allow exploration of physical conditions in 
the QGP prior to the onset of chemical reactions. Therefore, our
results for most peripheral reactions are   also 
somewhat   dependent on model assumptions about initial state and evolution
dynamics.

We have shows, in figures \ref{TwoVol}  and \ref{LHCVolVol2}, the impact parameter
dependence that arises in two volume expansion models.  Since  the analysis results
presented in Ref.~\cite{Rafelski:2004dp} have been used to fine tune 
the dynamical evolution model at RHIC, there is   good qualitative
agreement with experimental results. Our objective   has
been here to learn how to extrapolate the dynamics of strangeness 
production to the LHC environment. The gradual rise of 
strangeness yield $s/S$ with $dS/dy$, seen in figure \ref{sSdvdydhdy}, is  
reminiscent of the rise of $s/S$ with reaction energy obtained in a analysis of particle
yields obtained at SPS and RHIC at different reaction energies~\cite{Letessier:2005qe}.

We have made a (conservative) prediction regarding the increase in the K$^+/\pi^+$ 
ratio at LHC compared to RHIC, see figure \ref{LKpFLXOsS}. Though an important
result is that we expect an increase at LHC, we note that even a greater increase is
possible, signaling even greater values of $s/S$, depending on both:\\
1) the dynamics of the volume expansion --- this can
enhance the strangeness oversaturation of the final QGP state, 
see figures \ref{TwoVol} and \ref{LHCVolVol2};\\
2) QCD details, such as  strange quark density including
QCD interactions, and (still not well understood) strange quark mass,
see figures  \ref{alfdep} and \ref{massdep}.\\
In any case, we believe that this simple observable will show again
strangeness production growing faster than entropy production, 
its increase is directly coupled to an increase in $s/S$. We note again that in our
study the increase of $s/S$ with centrality implies that $\phi/{\rm K}^+$ also increases
with centrality. 

A natural result is the finding of the chemical yield equilibration of strangeness in 
the QGP formed in the 5\% most
central top RHIC energy reactions. This leads to
a better understanding of the resulting oversaturation of the hadron 
phase space by strangeness. The magnitude of this effect, 
dependent on the temperature of hadronization, can be considerable. 
This can be easily seen considering the magnitude of $s/S$ in both
QGP and HG phases. The final state hadrons 
formed far-off chemical equilibrium 
cannot significantly adjust    chemical composition, considering the 
rapid breakup of the fireball,  during the period 
of about  1--2 fm/$c$ prior to onset of the  free flow. Thus, our  finding is 
that strangeness rich QGP enhances  decisively the yields of 
multistrange hadrons. This phenomenon is more accentuated considering 
 charmed hadrons containing strangeness,
a topic under current investigation.
 
Furthermore, using the methods developed here, 
we have  considered thermal charm production. At LHC, we find
a nearly physically relevant thermal charm production, but not at RHIC. However, 
the thermal process we consider is able to produce enough charm 
 to   oversaturate the final state at both RHIC and LHC,
see figure \ref{Charm2Vol}. This also  shows that the direct parton
collision based production at RHIC  leads to extraordinarily large values 
of $\gamma_c$. The chemical nonequilibrium of charm is thus more pronounced 
than that of strangeness.

{\bf In conclusion:} The totality of our results shows
 that, as function of entropy yield $dS/dy$
(equivalently, the reaction energy of $A_1$--$A_2$ collision) and 
geometric reaction size (impact parameter dependence, 
participant number $A$), the phenomenon
of strangeness enhancement is well described by the mechanism of QGP based
thermal gluon fusion strangeness
production. We find both, as function of centrality, and energy,
a  gradual increase in specific strangeness yield, which agrees   with all
 available  experimental results. We find that, as function of energy, this 
continues  from  RHIC to LHC increasing our hopes for a more clear strangeness
signature of deconfinement.  

\vspace*{.2cm}
\subsubsection*{Acknowledgments}
Work supported by a grant from: the U.S. Department of Energy  DE-FG02-04ER4131.
LPTHE, Univ.\,Paris 6 et 7 is: Unit\'e mixte de Recherche du CNRS, UMR7589.
 

\vspace*{-0.3cm}

\end{document}